\documentclass[conference]{IEEEtran}
\IEEEoverridecommandlockouts
\usepackage{cite}
\usepackage{amsmath,amssymb,amsfonts}
\usepackage{algorithmic}
\usepackage{graphicx}
\usepackage{textcomp}
\usepackage{xcolor}
\usepackage{array}
\usepackage{amsmath}

\usepackage{multirow}
\usepackage{booktabs}
\usepackage{xurl}
\usepackage{subfigure}

\newcounter{KNNumberOfComments}
\stepcounter{KNNumberOfComments}

\newcounter{JSNumberOfComments}
\stepcounter{JSNumberOfComments}

\ifCLASSOPTIONcompsoc
\usepackage[caption=false,font=normalsize,labelfon
t=sf,textfont=sf]{subfig}
\else
\usepackage[caption=false,font=footnotesize]{subfi
g}
\fi

\def\BibTeX{{\rm B\kern-.05em{\sc i\kern-.025em b}\kern-.08em
    T\kern-.1667em\lower.7ex\hbox{E}\kern-.125emX}}
\begin{document}

\title{Characterizing Spontaneous Ideation Contest on Social Media: Case Study on the Name Change of Facebook to Meta}

\author{
\IEEEauthorblockN{Kunihiro Miyazaki}
\IEEEauthorblockA{\textit{The University of Tokyo}\\
Tokyo, Japan \\
kunihirom@acm.org}
\and
\IEEEauthorblockN{Takayuki Uchiba}
\IEEEauthorblockA{\textit{Sugakubunka}\\
Tokyo, Japan \\
takayuki.uchiba@sugakubunka.com}
\and
\IEEEauthorblockN{Haewoon Kwak, Jisun An}
\IEEEauthorblockA{
\textit{Singapore Management University}\\
Singapore\\
\{haewoon, jisun.an\}@acm.org}
}

\maketitle

\begin{abstract}
Collecting good ideas is vital for organizations, especially companies, to retain their competitiveness.
Social media is gathering attention as a place to extract ideas efficiently; however, the characteristics of ideas and the posters of ideas on social media are underexamined.
Thus, this study aims to characterize spontaneous ideation contests among social media users by taking an event of Facebook's name change to Meta as a case study.
As a dataset, we comprehensively collect tweets containing new acronyms of Big Tech companies, which we treat as an ``idea'' in this work.
In the analysis, we especially focus on the diversity of ideas, which would be the main reason for enlisting social media for idea generation.
As the main results, we discovered that social media users offered a wider range of ideas than those in mainstream media.
The follow-follower network of the users suggested that the users' position on the network is related to the preferred ideas.
Additionally, we discovered a link between the amount of user interaction on social media and the diversity of ideas.
This study would promote the use of social media as a part of open innovation and co-creation processes in the industry.
\end{abstract}



\section{Introduction}
Creating, developing, or communicating new ideas is an essential process for every organization. 
Since ideas are the source of creativity and innovation, much attention has been paid in academia and industry to developing and integrating them within and across organizations~\cite{boeddrich2004ideas}.
In an effort to collect good ideas, industries have been using the outsourcing of idea generation~\cite{gatzweiler2017dark}, especially in the form of \emph{ideation contest}. 
The ideation contest gathers participants and asks them for new ideas (e.g., about products or solutions in business) using crowdsourcing services and corporate platforms~\cite{bayus2013crowdsourcing,gamber2021effort}.
Companies such as Dell~\cite{bayus2013crowdsourcing}, IBM~\cite{Majchrzak2009HarnessingTP}, and Starbucks~\cite{muninger2019value} have succeeded in running the ideation contest to gather ideas and implement many of them. \looseness=-1

The great success in ideation contests has made companies explore the leverage of social media~\cite{piller2012social}.
Social media is a place where a wide variety of ideas are posted and discussed on a daily basis~\cite{sajid2016social}.
This big data would allow organizations to solicit ideas from a more expansive space and gain ideas from the types of people who would otherwise be excluded~\cite{jeppesen2010marginality}.
Due to its potential, existing research has proposed using social media for collecting ideas~\cite{bharati2021idea}; however, there is still a lack of understanding about how ideas are proposed on social media~\cite{cao2022visualizing,han2020computational}. \looseness=-1

\begin{figure}[!t]
\centering
\includegraphics[width=0.95\linewidth]{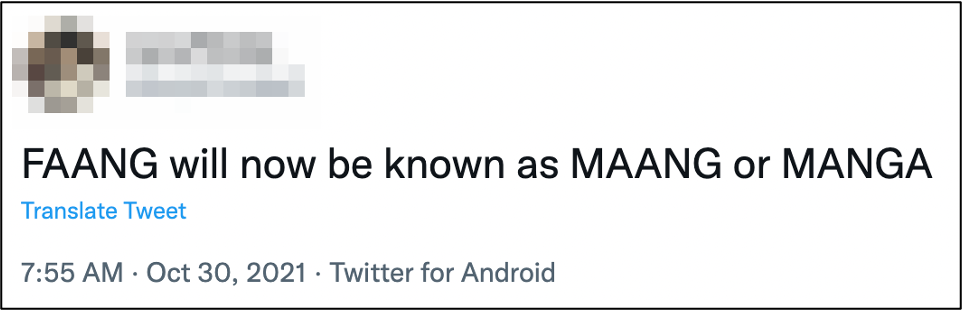}
\caption{Example of Idea tweet.}
\label{fig:example} 
\end{figure}

Therefore, this study characterizes idea generation behavior on social media. 
In particular, we analyze the diversity of ideas, an important concept when it comes to extracting ideas from social media.
As data, we deal with the acronyms of the American Big Tech companies, which have been known as FANG (Facebook, Amazon, Netflix, and Google) or FAANG (FANG + Apple)~\cite{FAANGSto80:online}. 
In October 2021, the name change of Facebook to Meta provided a unique opportunity to study a user-driven ideation contest on social media\textemdash  ``What should be a new acronym for Big Tech companies?'' 
Since the renaming of one of the world's top 10 companies by market capitalization was a phenomenal event, the discussion attracted many participants. 
In addition, an ideation contest about a name is referred to as a ``naming contest''~\cite{koh2019adopting}, which is categorized as the most straightforward form of ideation contests~\cite{yang2009open}.
Given these factors, Facebook's name change to Meta can serve as a solid exemplar of ideation contests on social media, and the empirical analysis of this case would provide helpful insights into the ideation behaviors of users. 
To characterize this spontaneous ideation contest, we exhaustively collect tweets in the discussion on new acronyms of Big Tech companies, which we consider as an ``idea'' in this work (an example is shown in Figure~\ref{fig:example}), and examine them.
After a first look at collected tweets, we conducted analyses on the diversity of ideas, the relationship between the users and ideas, and how user interaction affects the diversity of ideas.\looseness=-1


From the collected tweets, we found 21 candidate acronyms for referring to new Big Tech companies.
In the analyses of the diversity of ideas, we found that users on social media proposed a greater variety of ideas than those on mainstream media.
The user's follow-follower network indicated that network position is an important factor in affecting their ideation behavior.
Furthermore, we found that the interactions between users on social media are associated with a greater diversity of ideas.

Other than the analyses of the diversity of ideas, we characterized them in terms of the various metrics (e.g., shares, likes) and temporal dynamics.
Also, by looking at participants in detail, we found that the first person to post ideas tends to have fewer followers, while those with more followers post ideas secondarily or share the first posts. 
Finally, we conducted a regression analysis and found that the early participants of the users with more followers and negative posts were associated with more replies.


As far as we know, this work is the first case study to characterize spontaneous ideation contests on social media, and we obtained helpful insights for using social media for idea generation.
Data and code will be published when accepted and published.
We believe this study would promote the use of social media as a part of open innovation, co-creation processes, and the management of big data in the industry.

\section{Related Works}

\subsection{Ideation Contest using Crowdsourcing and Corporate Platforms}
To establish an efficient idea generation method, industries tried to study the mechanisms of idea generation in individuals~\cite{deo2021idea}, compare groups and individuals to determine which produce superior ideas~\cite{girotra2010idea}, and create an organizational system that facilitates idea generation from employees~\cite{van2020idea}.
However, with the increasing speed of information sharing about products, competition in business is intensifying every year, and thus the demand for collecting good ideas is increasing further~\cite{bjork2009good}.
In this context, companies are increasingly outsourcing idea generation~\cite{gatzweiler2017dark}.
Incorporating external knowledge, such as open innovation~\cite{huizingh2011open} or co-creation~\cite{ind2013meanings}, is expected to accelerate internal innovation and reduce the risk of product failure as it helps to involve potential customers in idea generation~\cite{von2016free}.
Among the means to outsource external ideas, ideation contest using crowdsourcing services~\cite{bayus2013crowdsourcing} and corporate ideation platforms~\cite{gamber2021effort}, which are mainly financially incentivized~\cite{ihl2012all}, is one of the leading choices~\cite{piller2012social,gatzweiler2017dark}.

Existing research summarized the types of ideation contests in addition to naming contests, such as graphic design and creative writing contests~\cite{yang2009open}.
As for the analysis on the corporate ideation platform, Bayus analyzed their behavior and the quality of their ideas with respect to serial ideators and ideators with only one idea~\cite{bayus2013crowdsourcing}. 
Hossain et al. analyzed the growth process of the platform and the relationship between the total number of ideas and the number of viable ideas~\cite{hossain2015ideation}.
In contrast, we analyze spontaneous ideation contests on social media, which have the advantage of obtaining various participants' opinions with less bias than corporate ideation platforms~\cite{jeppesen2010marginality}. \looseness=-1

\subsection{Ideation Contest using Social Media}
The problem with ideation contests with crowdsourcing services and corporate platforms is that the number of participants is limited and sometimes biased~\cite{piller2012social}.
In this context, the use of social media for ideation contests is gaining attention to supplement existing methods~\cite{piller2012social}.
Social media is rich in various kinds of ideas, and there is a strong demand for methods to use these ideas in corporate innovation.
Due to these properties, researchers have proposed the inclusion of social media information in the business innovation process~\cite{piller2012social} and demonstrated its benefits theoretically~\cite{bharati2021idea}.
In fact, many companies attempt to use social media to enhance their innovation process in some forms~\cite{roberts2016finding}, and idea creators such as fashion designers have been using social media frequently in recent years to encounter new inspiration~\cite{banks2014social,villioti2018social}.\looseness=-1

So far, the primary use of social media in the industry has been to look up related words or feedback for a specific product as marketing research~\cite{cole2015social}, and there has been little study on the mechanism of how social media users generate ideas.
As the close studies, Carr et al. searched for the related words of a product from social media, analyzed their volume, and confirmed the insights obtained the result aligned with existing research methods~\cite{carr2015social}. 
Han et al. held a design challenge in which designers were asked to design a chair based on the chair's related words and feelings obtained through social media searches and found that social media information is helpful for designers~\cite{han2020computational}.
Ozcan et al. tried to make a model to classify whether a post on social media contains ideas or not~\cite{ozcan2021social}.
Our study advances this line of research and is the first empirical analysis of an ideation contest on social media. \looseness=-1

Other papers examined hashtags and their evolution on social media with respect to what words are used for certain events.
Cunha et al. characterized what hashtags are used in real-world events and showed the structure of extreme usage of some hashtags~\cite{cunha2011analyzing}.
Sato et al. visualized how hashtags evolve over time in a network with edit distance between hashtags as edges, and showed that hashtags are evolving from seeded hashtags in various directions over time\cite{sato2021exploration}.
On the other hand, these studies differ from ours in that they analyzed various hashtags describing a specific event together, which included their impressions of the event, and not for the purpose of ideation for naming purposes.
Furthermore, these papers focus solely on hashtags, and our study further analyzes the posters of ideas and their interaction.

\subsection{Collective Creativity and Diversity of Ideas}
Idea generation is an essential component of the production process and one of the highest leverage points for a company~\cite{dahan2002product}.
For an efficient idea generation, the exchange of ideas by many individuals is considered effective~\cite{janis1977decision} because it stimulates further associations and ideas~\cite{paulus2014creativity} and provides exposure to more ideas than generating ideas alone~\cite{csikszentmihalyi1997flow}.
This kind of approach to creative activity is called collective creativity~\cite{inakage2007collective}.

Collective creativity is sometimes criticized as being ineffective for idea generation.
For example, some studies show that brainstorming does not efficiently aid in the generation of ideas~\cite{west1996innovation}.
However, the underlying argument is that the evaluation of submitted ideas on the spot prevents the submission of the next idea~\cite{leenders2003virtuality}, and it has been reported that this drawback can be overcome by, for example, exchanging ideas in a text-based manner instead of gathering ideas in-person~\cite{dennis1993computer}.
Furthermore, unlike brainstorming, social media has the potential to extract more diverse ideas because it is an environment for casual idea generation, where submitted ideas do not necessarily get feedback from others.

In collective creativity, the key to sound idea generation is the diversity of participants and ideas~\cite{johnson2021neglect}.
Innovation is said to come from the combination of knowledge from different fields~\cite{parjanen2010collective,parjanen2012brokerage}; thus, the stimulus supplied to the participants in the exchange of ideas would increase with the diversity of the participants and the ideas, which results in the generation of more fresh ideas.
It was shown in laboratory experiments~\cite{parjanen2012brokerage} that mediating distant people (i.e., weak connections) in a social network can facilitate the generation of new ideas.
Considering that social media has a higher contingency for far-distant people on social networks than in the physical world and that it is easier to mediate distant users, social media is a promising place for generating new ideas.

Despite the potential of social media to extract diverse ideas, there is still a lack of understanding about how to extract them efficiently.
By analyzing ideation behaviors in social media, this study confirms the potential for utilizing them for organizational open-innovation and explores how to utilize them.


\section{Data Collection}


\subsection{Finding Candidate Acronyms}

We first collect tweets to identify candidate acronyms.
We use the Twitter Academic API~\cite{Enabling63:online} to obtain tweets containing ``FANG'' or ``FAANG'', which have been common names for Big Tech~\cite{FAANGSto80:online}, from 28 Oct 2021 to 30 Nov 2021, one month window after Facebook's name change.
The API is case-insensitive in search.
As a result, we obtained 53,975 English tweets without retweets (RTs).
Then, after making all text lowercase, we extracted the 150 most co-occurred words with  ``fang'' or ``faang'' based on the Jaccard coefficient~\cite{niwattanakul2013using}, and conducted a manual examination to extract a set of candidate acronyms of the new Big Tech companies.
As a result, we obtained 21 candidate ideas (shown in Table~\ref{table:candidates}). 
Note that we omitted the candidates with a single tweet, which are MAMSANG, TMAANG, MAANAT, MANATAM, and MATANTA, from Table~\ref{table:candidates}.

\begin{table}[!htb]
\centering
\caption{The  candidate ideas and their statistics. }
\resizebox{9.0cm}{!}{
\begin{tabular}{lrccccrc}
\hline
Idea    & Tweet & RT    & Like  & Reply & QT   & UU  & Follower \\ \hline
MAANG   & 885   & 0.62 & 5.70  & 0.97  & 0.19 & 850 & 93151.9  \\
MANGA   & 793   & 1.55 & 11.28 & 0.93  & 0.22 & 759 & 18760.1  \\
MANG    & 135   & 1.01 & 5.41  & 0.93  & 0.37 & 133 & 5901.5   \\
MAMAA   & 132   & 0.55 & 2.42  & 0.61  & 0.17 & 129 & 43564.4  \\
MAANA   & 115   & 0.60 & 6.92  & 0.56  & 0.17 & 111 & 5013.6   \\
MAGMA   & 89    & 1.06 & 12.99 & 1.73  & 0.33 & 83  & 2692.2   \\
GAMMA   & 87    & 0.34 & 5.68  & 0.98  & 0.09 & 84  & 3386.5   \\
MAGA    & 44    & 1.66 & 31.45 & 1.95  & 0.52 & 44  & 3668.2   \\
TAANG   & 23    & 0.04 & 1.65  & 0.39  & 0.04 & 21  & 1337.6   \\
MAMA    & 17    & 0.29 & 3.71  & 0.35  & 0.12 & 17  & 3224.3   \\
MAGNA   & 16    & 0.00 & 1.06  & 0.38  & 0.06 & 16  & 2097.8   \\
MAAMA   & 15    & 0.73 & 10.13 & 1.27  & 0.13 & 15  & 6493.3   \\
MAMATA  & 13    & 6.92 & 58.38 & 4.46  & 1.23 & 13  & 185656.1 \\
MAMANG  & 8     & 0.13 & 1.00  & 0.75  & 0.00 & 8   & 1768.1   \\
AMAMA   & 7     & 2.29 & 6.71  & 0.43  & 0.14 & 7   & 14150.4  \\
MAANAM  & 3     & 0.33 & 8.00  & 0.33  & 0.33 & 2   & 1662.0   \\
\hline
\end{tabular}
}
\label{table:candidates}
\end{table}


\subsection{Collecting Tweets with Candidate Acronyms}
We use the Twitter API again to search for tweets containing those 21 candidates over the same period.
As a result, we retrieved 3.55 million English tweets, excluding RTs.
However, we found that many of the candidates had homonyms (e.g., MAMA for a music award~\cite{MnetAsia52:online}, MAGA for a political slogan~\cite{MakeAmer35:online}). 
Therefore, to extract relevant tweets for this study, we focus on the tweets that match either of the following two conditions:
\begin{itemize}
    \item Condition 1: Tweets that contain (FANG, FAANG, or Meta) AND (any one or more of the candidate words);
    \item Condition 2: Tweets that contain (any one or more of the candidate words) AND (all the company names in the corresponding candidate word).
\end{itemize}
An example of condition 2 is a tweet containing `MANG,' AND `Meta,' `Apple,' `Netflix,' and `Google.'
As some companies could represent the same acronym (e.g., `Amazon' and `Apple' could be used for `A' in an acronym), we used all the combinations of companies for searching the corresponding tweets using the possible company names listed in $\S$3.3.
We consider both lowercase and uppercase for search.  
Then, we checked all the tweets manually and excluded 74 tweets that had nothing to do with the context of Big Tech, which were mainly composed of unremoved MAMA tweets~\cite{MnetAsia52:online} and hashtag hijacking~\cite{xanthopoulos2016hashtag}.
As a result, we retrieved 2,219 tweets (Idea tweets) excluding RTs, of which 1,766 were regular tweets that were neither replies nor quote retweets (QTs). 
A drastic decrease in the number of tweets is mainly due to the massive amount of tweets related to ``MAMA,'' 1.6 million tweets in the original dataset.
Our filtering criteria were found to be effective in removing irrelevant tweets; even among the tweets containing MAANG, we filtered out 38.8\% of tweets that are unrelated to Big Tech (e.g., MAANG seems to be a casual word used often in Indonesian users). 
We then collected how other users engaged with these Idea tweets: 3,735 replies, 850 QTs, and 1,622 RTs.

We also collected the followers of (1) the authors of Idea tweets and (2) those who share (by RT or QT) the Idea tweets.
The number of users is 3,912. 
Out of the number of followers of each user, the max is approx. 24.2 mil., the minimum is 1, the mean is 18,117.0, and the median is 366.5.

\section{A First Look at the Spontaneous Ideation Contest}

\subsection{A List of Ideas and User Engagement Statistics}
The ideas (i.e., candidate acronyms), the frequency of tweets containing them, and various engagement metrics are shown in Table~\ref{table:candidates}, which include the number of posts containing them (Tweet), retweet (RT), like, reply, quote retweets (QT), unique users (UU), and the follower counts of the authors of Tweet (Follower). 
The metrics except for Tweet and UU indicate the average values per tweet.\looseness=-1 

All the candidates are acronyms of A (Apple, Amazon, or Alphabet), G (Google), M (Microsoft or Meta), N (Netflix), S (Salesforce), and T (Tesla or Twitter). 
Interestingly, compared to the original FAANG companies, several companies are newly included in candidate acronyms, which might reflect the dynamics of the IT industry. \looseness=-1 

In terms of the number of tweets, MAANG and MANGA are the two most popular ideas.
MAANG is a simple replacement of F (Facebook) with M (Meta) in FAANG.
MANGA is one of the famous Japanese words that refer to Japanese comics~\cite{MangaWik51:online}. 
We suppose that the popularity of MANGA is due to the serendipitous match with the well-known term. 
MANG, which is also a direct replacement for FANG, is the third most popular idea, presumably because FAANG has been more widely used than FANG since 2017~\cite{FANG_FAANG}.
In fact, FANG (141 tweets) was much less common than FAANG (1,766 tweets).\looseness=-1 

For other engagement metrics, different ideas are more popular than MAANG and MANGA: MAMA, MAGA, and MAMATA for RTs and likes\textemdash more users are engaged by clicks; MAGMA, MAGA, and MAMATA for the reply and QT\textemdash more user are engaged by leaving a tweet with their own text; MAMATA and AMAMA for the average number of followers\textemdash more influencers are engaged.
The number of Idea tweets and their unique users is almost the same, indicating that each user tweeted an idea once.\looseness=-1 

We note that 150 (8.77\%) out of the 2,219 Idea tweets contained multiple ideas, with the number of simultaneous ideas where the max is 7, the minimum is 2, the mean is 2.12, and the median is 2.
MAANG and MANGA appeared simultaneously in the most Idea tweets (60 times), while the combination of MAANA and MANG had the highest Jaccard coefficient (0.09).\looseness=-1

\subsection{Time Series of the Ideation Contest}

\begin{figure}[!htb]
\centering
    \subfigure[Tweet]{
        \includegraphics[width=0.48\linewidth]{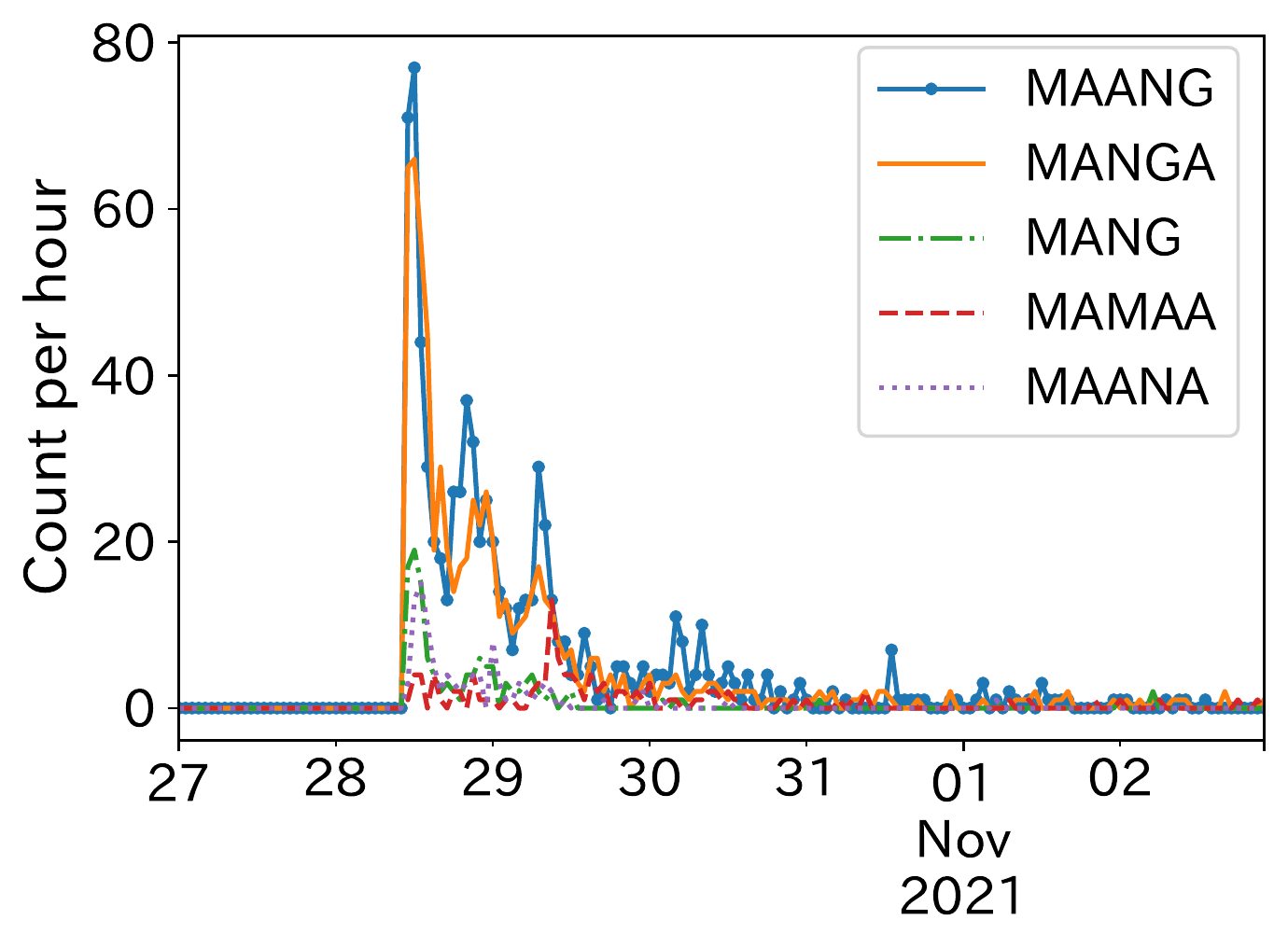}
    }  
    \subfigure[Retweet]{
        \includegraphics[width=0.48\linewidth]{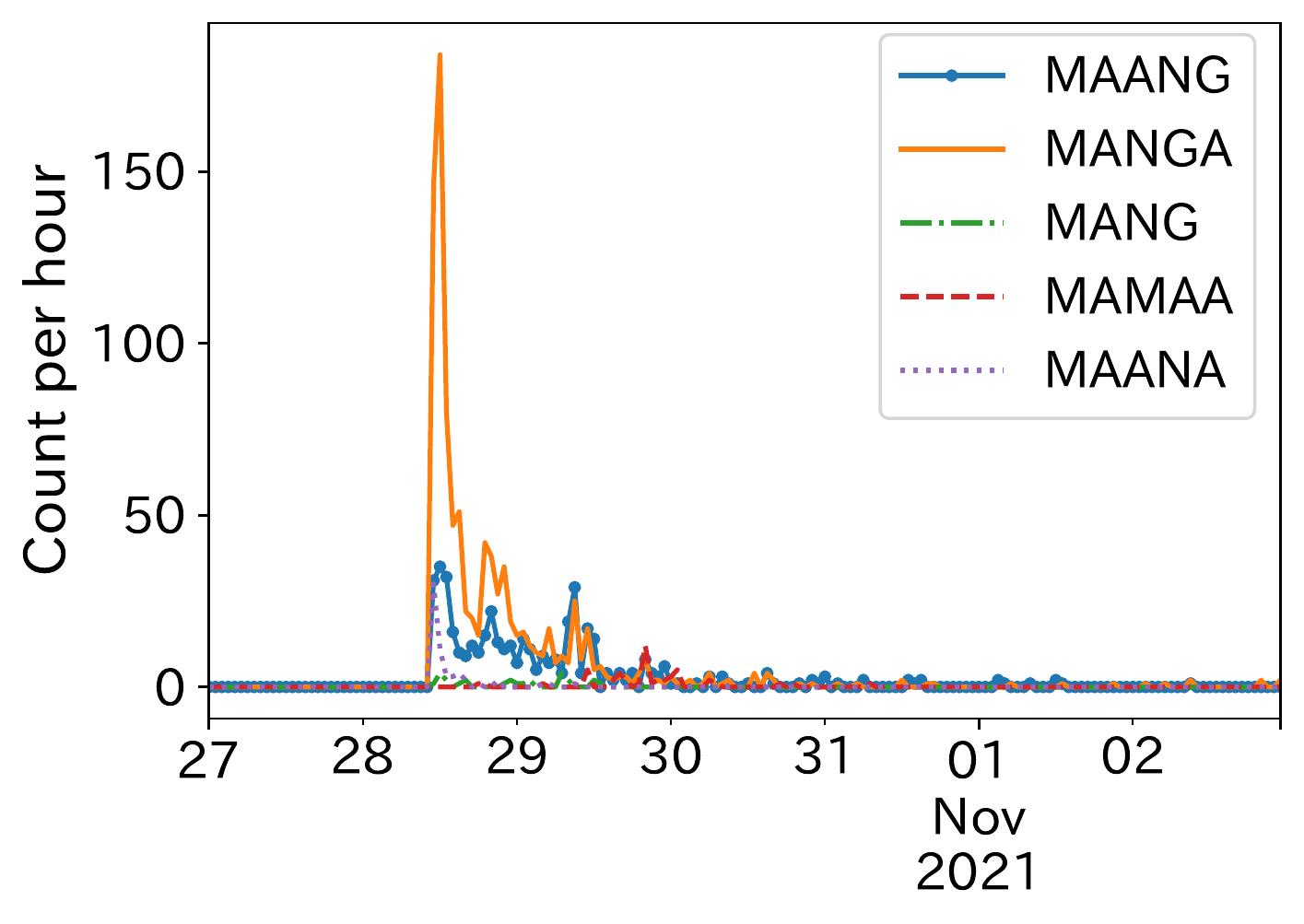}
    }
\caption{Hourly frequency of tweets and retweets of top five ideas.}
\label{fig:temporal_hourly}
\end{figure}

To investigate how the ideation contest rises and converges, we examine the time series of the idea tweets.
Figure~\ref{fig:temporal_hourly} presents the hourly volume of tweets with the top 5 ideas and the hourly volume of RTs of those tweets.
The period is one week, starting from the day before  Facebook changes its name to Meta.
We can see that all the ideas emerge at almost the same time, which is right after the announcement of Facebook's name change.
When looking into the dynamics on a finer scale, we found that the tweet of a news account that first reported about Facebook's name change was written at  6:07 PM on October 29, 2021 (UTC+0)~\cite{Fintwito76:online} as far as we could confirm. 
The ideation contest started around 16 minutes after the name change announcement.\looseness=-1 

The number of ideas gradually increased after the announcement, with convergence and excitement repeating several times. 
By October 31, three days after the announcement, the contest is almost over.
When looking at the volume of tweets for each candidate over time, it is hard to judge which one is dominant between MAANG and MANGA at the start, and the gap is gradually opening up. 
Also, in terms of RTs, MANGA is overwhelmingly large initially, but in line with MANGA's gradual dominance in the number of tweets, MANGA has been overtaken by MAANG in hourly RTs at the third peak. \looseness=-1 

\section{Diversity of ideas on social media}
\subsection{Comparison with Mainstream Media}
As we are especially interested in how diverse ideas are proposed on social media, an analysis of the ideas mentioned in mainstream media and media-related accounts can provide a good reference.  

To this aim, we divide the users who post Idea tweets into media-related accounts and the rest.
For this purpose, we extracted 159 users whose account names and bio information include the words ``media,'' ``news,'' ``reporter,'' ``journalist'' among 2,081 users we collected.
Then, we manually remove accounts that are not related to the media from them.
In addition, to mark prominent media, e.g., CNBC and Reuters, we use Twitter's ``verified'' information associated with the account~\cite{TwitterV40:online}.
As a result, we obtained 93 media-related (of which verified: 17) and 1,988 non-media-related users (of which verified: 62). \looseness=-1

\begin{figure}[!htb]
\centering
\includegraphics[width=0.75\linewidth]{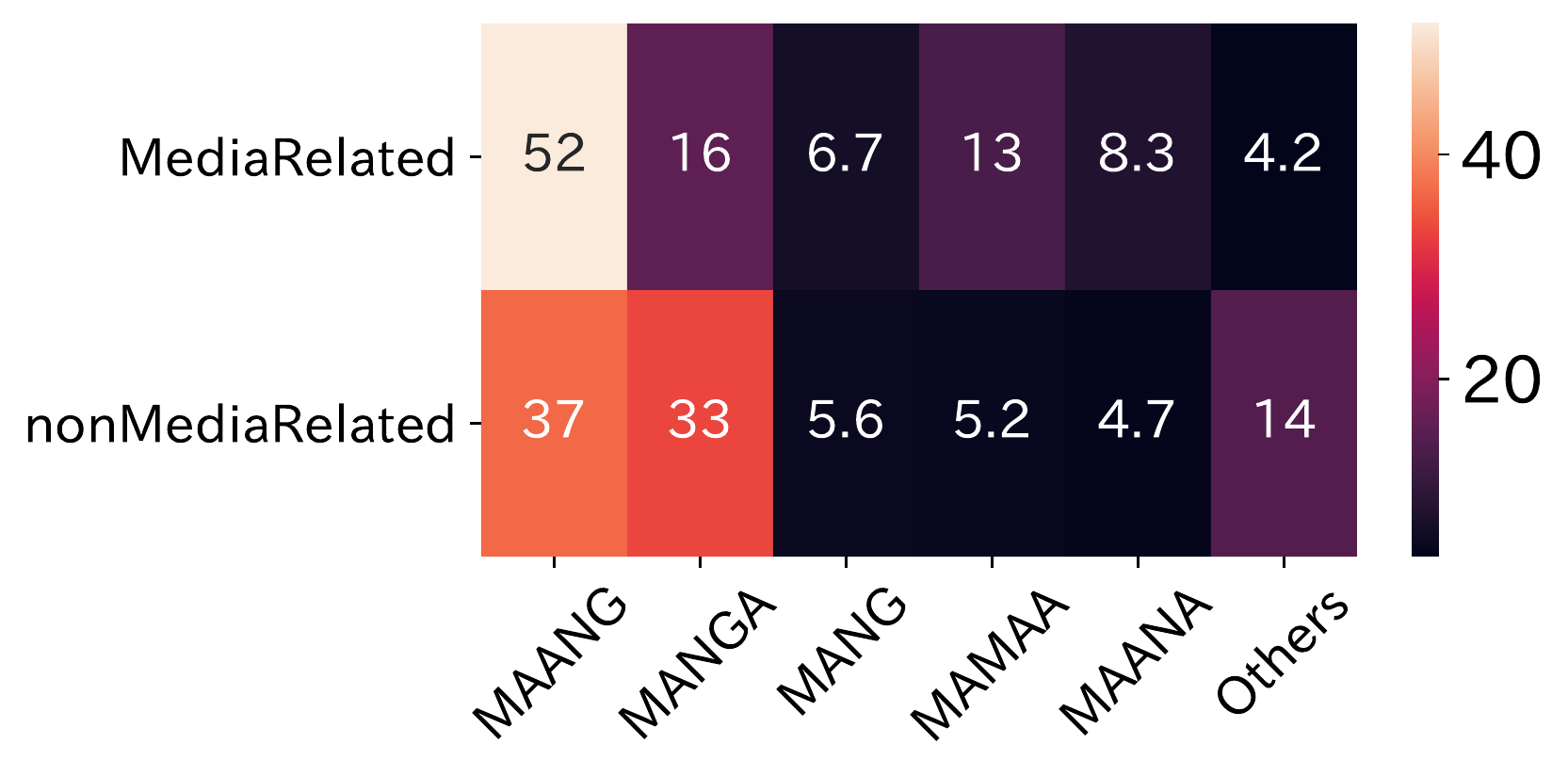}
\caption{
Relationship between (non-)media-related accounts and ideas. 
The numbers (aligned with colors) indicate the percentage of ideas posted by each group of users (the sum of row numbers is 100). 
}
\label{fig:heatmap_Media} 
\end{figure}

\begin{figure}[!htb]
\centering
\includegraphics[width=0.97\linewidth]{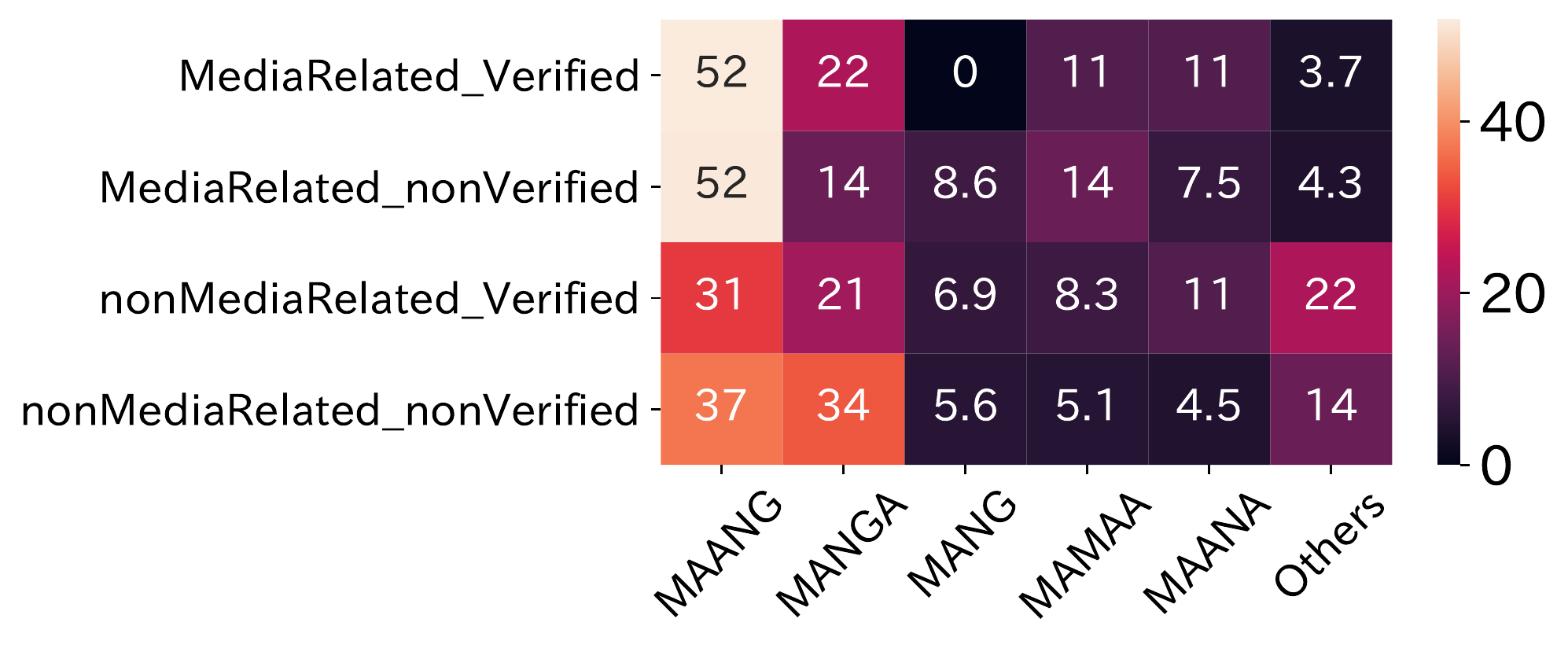}
\caption{
Relationship between (non-)media-related accounts and ideas. 
The accounts are further classified by ``verified'' information. 
The numbers (aligned with colors) indicate the percentage of ideas posted by each group of users (the sum of row numbers is 100).
}
\label{fig:heatmap_Media_verified} 
\end{figure}

The distribution of the ideas they mention is shown in Figure~\ref{fig:heatmap_Media} and \ref{fig:heatmap_Media_verified}. 
These figures show the counts of Idea tweets aggregated by each idea and converted into percentages; here, the top five ideas and the aggregation of other ideas (Others) are presented.
Comparing media-related accounts with the rest (Figure~\ref{fig:heatmap_Media}), we find that media-related accounts are more heavily biased toward MAANG, while there is a much less ratio for Others, meaning a biased distribution of ideas. 
The $\chi^2$ tests showed a significant difference in the distribution of ideas between media-related and non-media-related accounts ($p < 0.001$). 
Note that we conduct the $\chi^2$ tests on count-based aggregation, not on ratios~\cite{ChiSquar48:online}, and the same for the later analyses.
Furthermore, in Figure~\ref{fig:heatmap_Media_verified}, we see that both verified and non-verified media-related accounts exhibit skewed distributions of ideas (in both cases, the higher ratio for MAANG and the lower ratio for Others).
This suggests that a diversity of ideas are more likely to surface on social media than in the mainstream media, regardless of their prominence.\looseness=-1

\subsection{Comparison with Influencers}
\begin{figure}[!htb]
\centering
\includegraphics[width=0.75\linewidth]{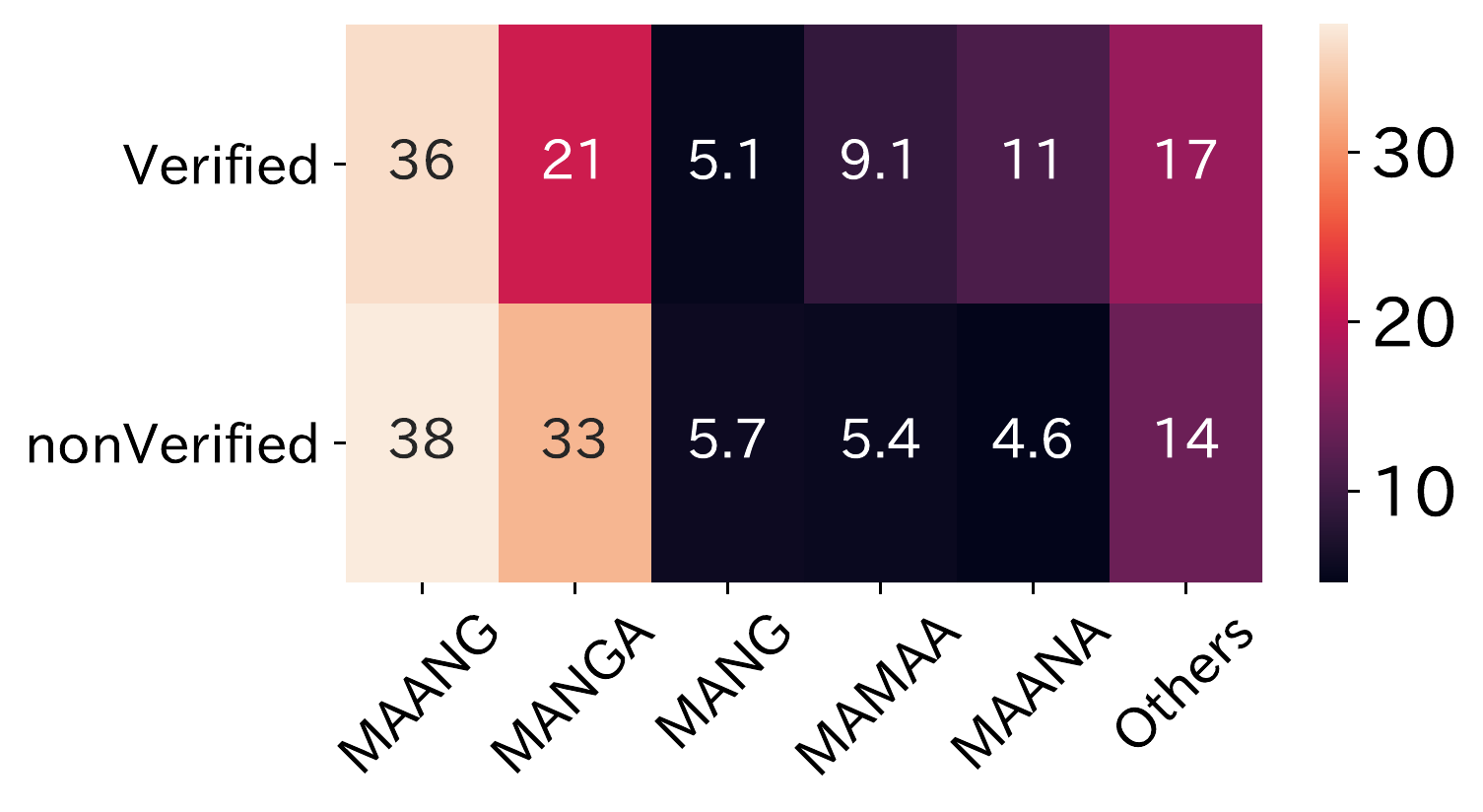}
\caption{
Relationship between (non-)verified accounts and ideas. The numbers (aligned with colors) indicate the percentage of ideas posted by each group of users (the sum of row numbers is 100).
}
\label{fig:heatmap_verified} 
\end{figure}

We also show the heatmap for (non-)verified accounts in Figure~\ref{fig:heatmap_verified}.
Interestingly, there is no significant difference between the two groups ($p = 0.33$ by the $\chi^2$ test). 
Rather, the distribution of ideas seems flatter for verified accounts (the Gini coefficients: 0.332 for the verified group and 0.425 for the non-verified group).
Intuitively, it seems that users with less presence (less influential) may enjoy more freedom to propose diverse ideas.
To validate this intuition quantitatively, we analyze the relationship between the number of followers of accounts and the diversity of ideas in this subsection.\looseness=-1 

First, we divided the users into four quadrants based on the number of followers; the first quartile is 73, the median is 303, and the third quartile is 1,322.
As the number of followers in the third quartile, 1,322, is relatively small to be called an influencer in the Twitterverse, we further divide the users by the number of followers by 10k (151 users) in reference to the discussion in \cite{Howtouse34:online}.

\begin{figure}[!htb]
\centering
\includegraphics[width=0.75\linewidth]{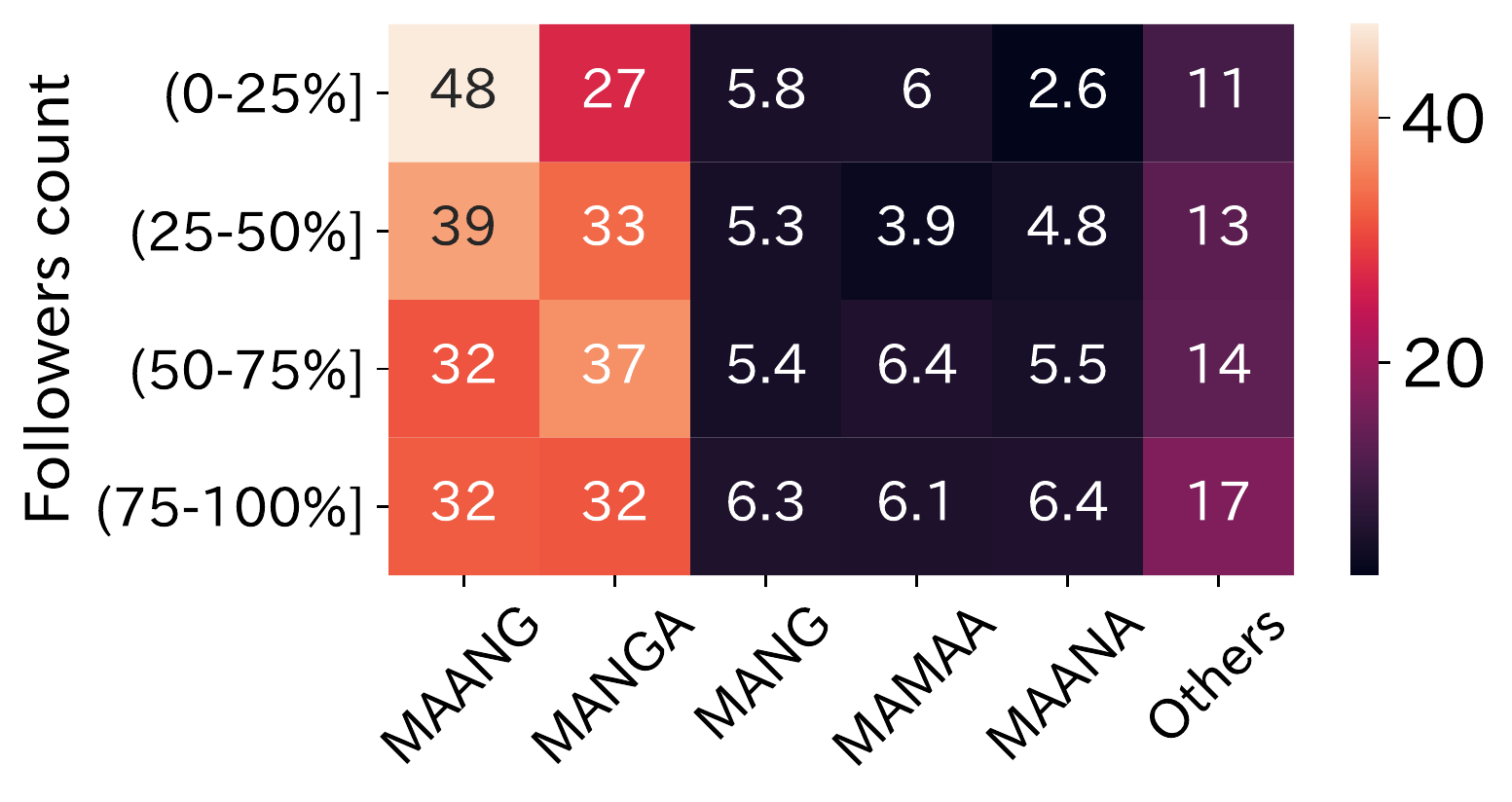}
\caption{
Relationship between followers count (by four quadrants) and ideas. The numbers (aligned with colors) indicate the percentage of ideas posted by each group of users (the sum of row numbers is 100).
}
\label{fig:heatmap_followers_quantilecut} 
\end{figure}

\begin{figure}[!htb]
\centering
\includegraphics[width=0.75\linewidth]{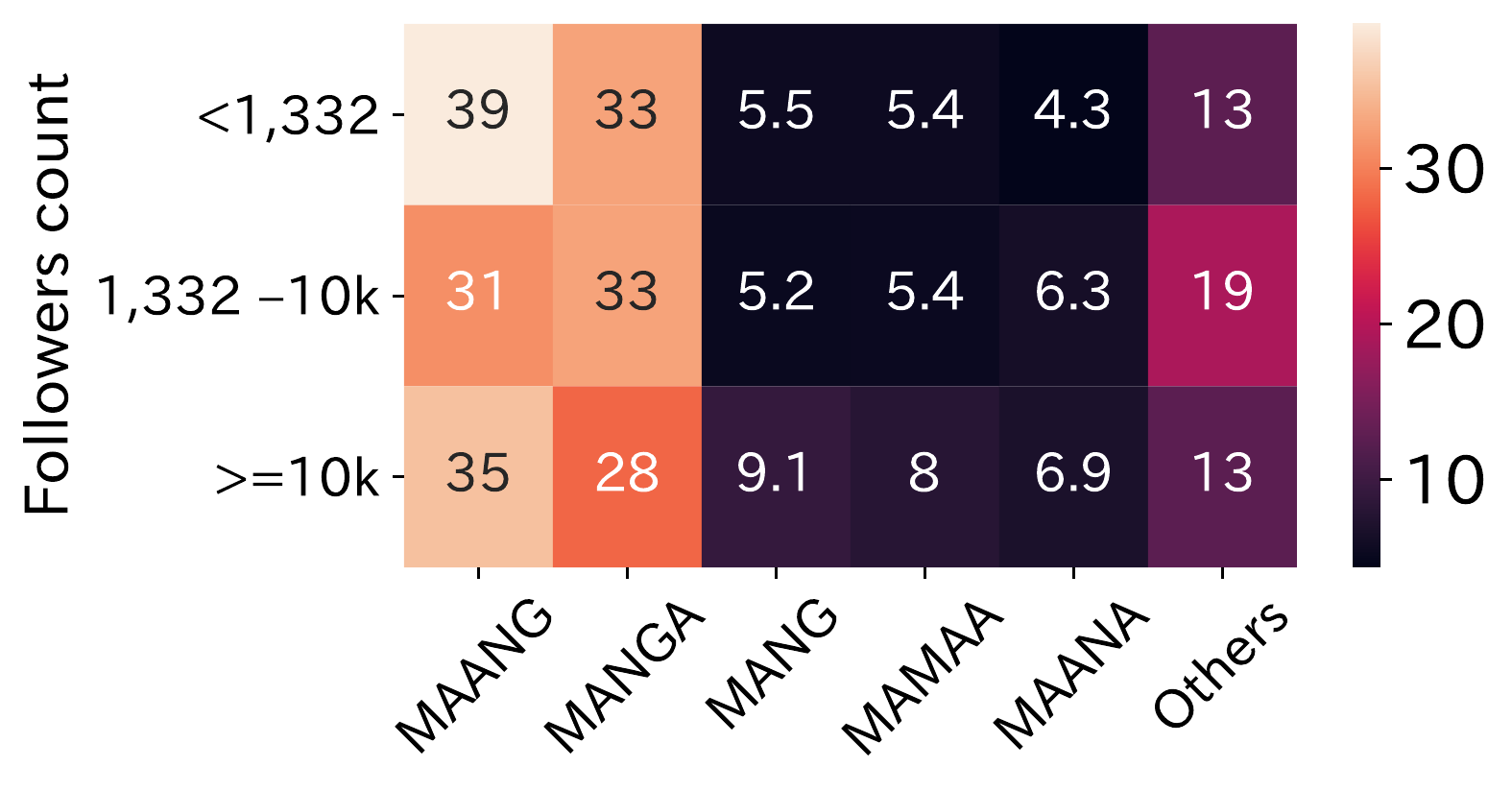}
\caption{
Relationship between followers count (by the third quartile (1,322) and 10k) and ideas. The numbers (aligned with colors) indicate the percentage of ideas posted by each group of users (the sum of row numbers is 100).
}
\label{fig:heatmap_followers_10000} 
\end{figure}

Figure~\ref{fig:heatmap_followers_quantilecut} and \ref{fig:heatmap_followers_10000} show the heatmaps by the number of followers.
Interestingly, in Figure~\ref{fig:heatmap_followers_quantilecut}, as the number of followers increases, the proportion of MAANG decreases and the proportion of Others increases.
In other words, the larger number of followers the users have, the more diverse their ideas are.
This result is somewhat unexpected because it was expected from past studies that users are less likely to propose ideas when they are evaluated by others~\cite{leenders2003virtuality}. 
We assume they have a large number of followers because they usually post interesting ideas.\looseness=-1 

On the other hand, Figure~\ref{fig:heatmap_followers_10000} shows that users in the range from the third quartile (1,322) to 10k, or middle influencers, generate the most diverse ideas, rather than users with more than 10k followers (although $p = 0.10$ for the $\chi^2$ test of `1,332 –10k' between `$>=$10k'). 
Thus, there is the possibility that the aforementioned study~\cite{leenders2003virtuality} is only applicable to the users with more than 10k followers: when users have a high probability of getting feedback on their ideas, it is more difficult to propose unique ideas because of their publicity. \looseness=-1 

\section{Network and Profiles of Participants}

\subsection{User Groups and Preferences of Ideas}

\begin{figure*}[!ht]
\centering
\includegraphics[width=0.90\linewidth]{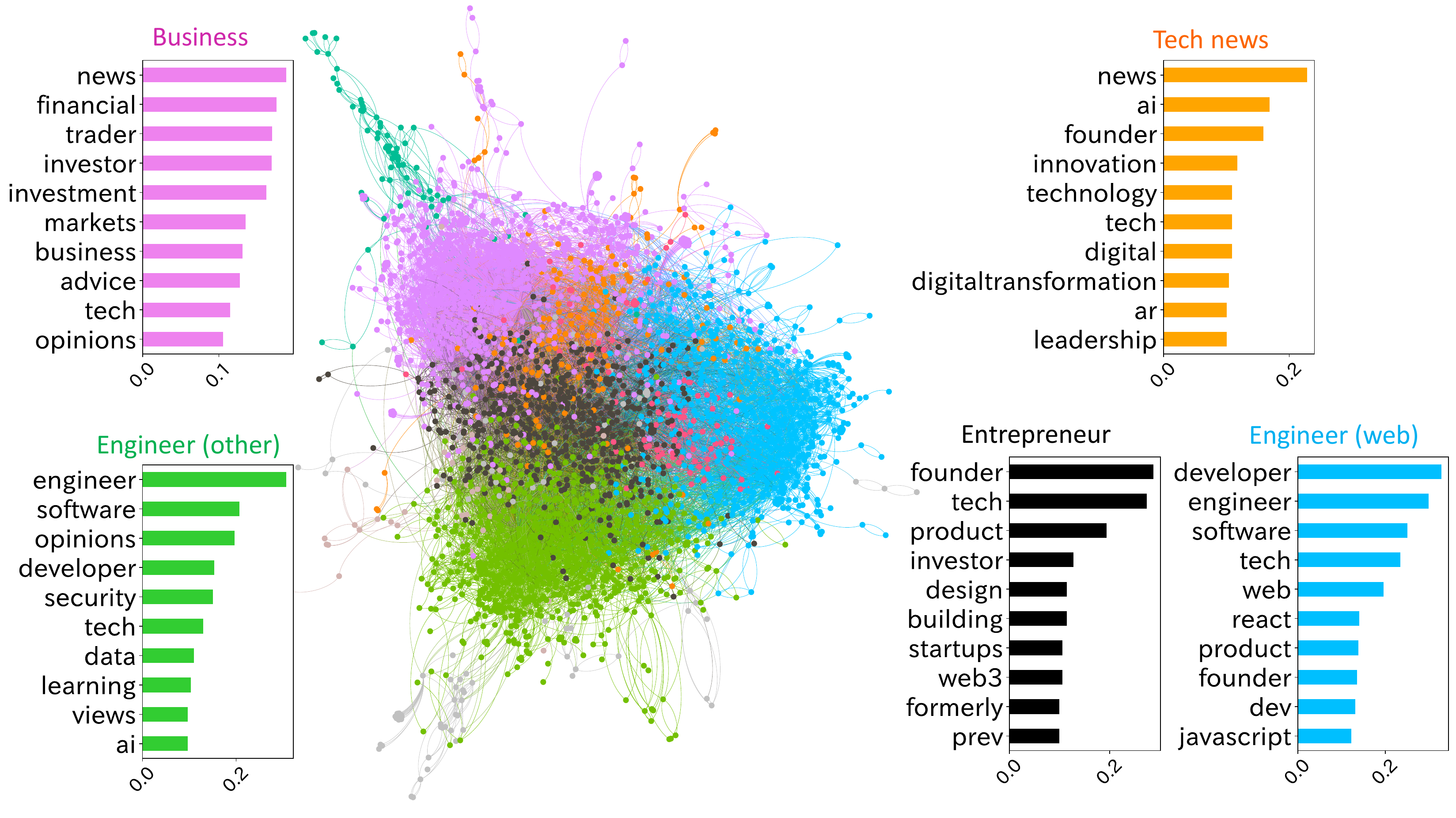}
\caption{
The largest connected component of the follower network of accounts that proposed and shared ideas. 
Nodes represent users and edges represent follow-follower relationships (undirected), and colors correspond to the clusters by Louvain method~\cite{blondel2008fast}. 
The representative words of bio information and TF-IDF scores in each cluster are also displayed with the colors corresponding to the color of clusters in the network.
}
\label{fig:follower_network} 
\end{figure*}

Now we move on to \emph{who} proposes and spreads the ideas. 
Existing research suggests that there is a relationship between ``distance of participants'' and diversity of ideas, with more distant people having different ideas~\cite{parjanen2012brokerage}. 
On social media, which is a virtual environment, distance is often measured from a social network constructed by the users' relationships~\cite{backstrom2012four}.
In this section, we study the relationship between the distance of users on social media and their idea preferences.\looseness=-1 

To this aim, we collect and reconstruct the following network of users who propose and spread ideas.
We found that most accounts were connected, with the ratio of its largest connected component as 79.7\% of all users.
On the other hand, the other connected components consisted of less than three users.
To examine different groups joining in the idea contest, we identified clusters of the largest connected component by using Louvain clustering~\cite{blondel2008fast} (with the ``resolution'' parameter as 1), resulting in the five largest clusters in the network. 
Figure~\ref{fig:follower_network} visualizes the network colored by different clusters with their representative words. 
The shape of the network is formed by the ForceAtlas2 algorithm~\cite{jacomy2014forceatlas2}.
We determined the representative words for each cluster by extracting the words with the highest TF-IDF scores in the users' bio information of each cluster. 
When calculating TF-IDF scores, we aggregated the texts in user bios into one document for each cluster (i.e., we made five documents for the calculation). 
By manually looking at the representative words, we labeled each cluster as Business, Engineer (web), Engineer (other), Entrepreneur, and Tech news. Each cluster accounts for 28.4\%, 19.2\%, 22.2\%, 15.9\%, and 4.8\% of the largest connected component, respectively. 
Seemingly, the clusters are largely composed of tech and business-related accounts.
Their high engagement in the ideation contest makes sense because 1) Big Tech companies are deeply related to engineers in terms of their products and careers, and 2) since these companies are significant investment targets, they drew interest from business and investment accounts.
We note that we manually examined that the Business cluster has many individuals (not news organizations) who had the word ``news'' in their bio, but the Tech news cluster has many news organizations. \looseness=-1

\begin{figure}[!htb]
\centering
\includegraphics[width=0.90\linewidth]{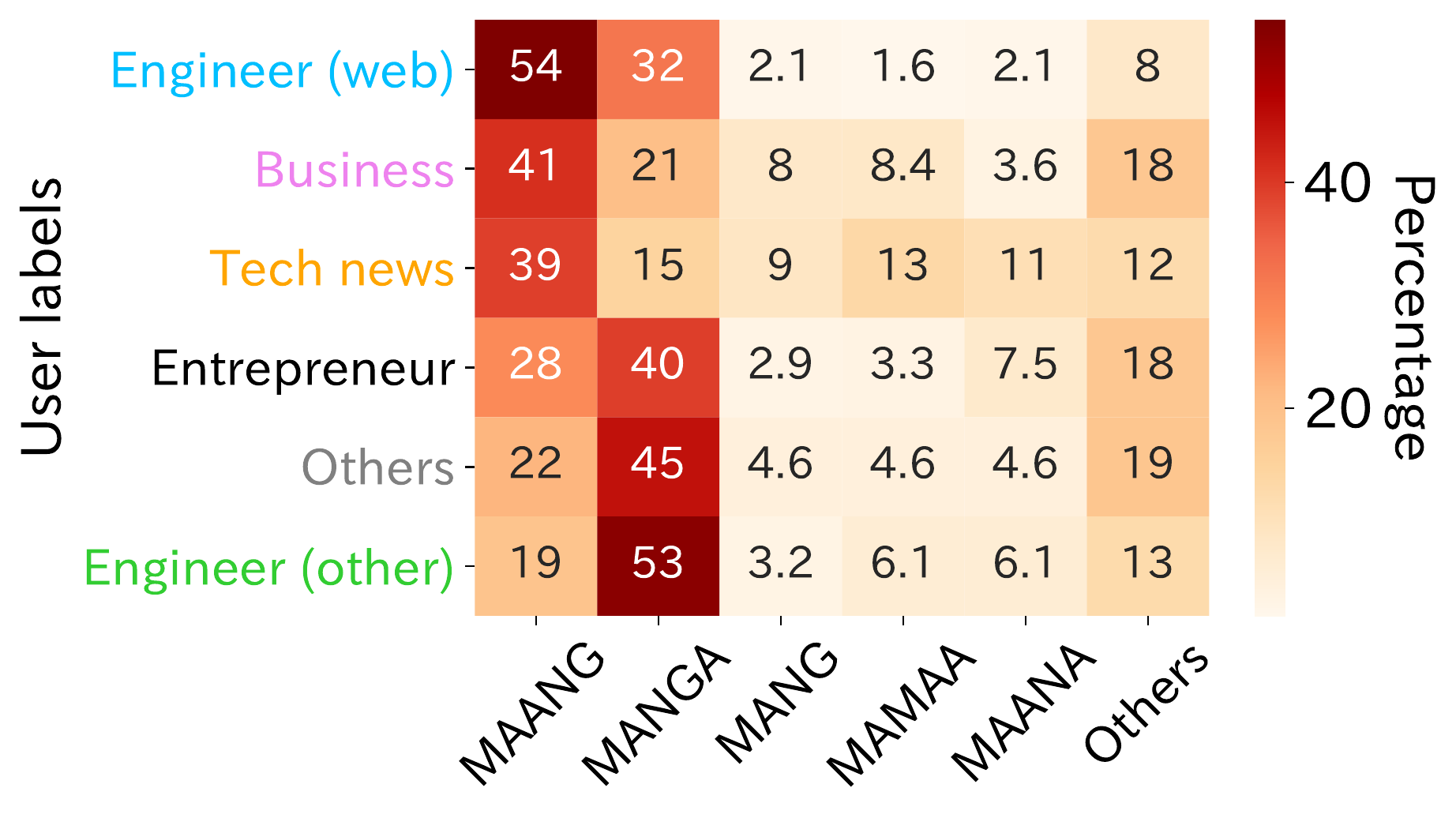}
\caption{
Relationship between user clusters and ideas. The numbers (aligned with colors) indicate the percentage of ideas posted by each group of users (the sum of row numbers is 100).
The rows are sorted by the percentage for MAANG.
}
\label{fig:heatmap2} 
\end{figure}

Then, how do these clusters relate to the ideas? 
Figure~\ref{fig:heatmap2} is a heatmap showing the ratios of ideas for the above-mentioned top 5 groups and the aggregation of users in the rest of the groups (Others).
The pair-wise $\chi^2$ tests showed a significant difference in the distribution of ideas in all pairs (all $p < 0.001$) except for the pair of `Engineer (other)' and `Others' ($p = 0.49$) and the pair of `Entrepreneur' and `Others' ($p = 0.57$).
In other words, this result provides evidence that distant users on social media prefer different ideas, which aligns with the previous research~\cite{backstrom2012four}.\looseness=-1 

This result can also come from the factor that people with different jobs prefer different ideas.
Given that people with similar attributes tend to cluster close on social networks~\cite{bisgin2012study}, it is natural to assume a relationship between these occupations and distance.
Also, note that the present analysis does not determine which affects the diversity of ideas more, distance or attributes. 
Nor does the result indicate a causal relationship: the result does not indicate that distancing users or changing attributes of users will necessarily result in a change in their preferred ideas.
However, it is possible that distance leads to a preference for different ideas, even if the attributes are similar. 
This is because the present network shows that the engineers in different locations significantly differed in the preferred ideas (MAANG and MANGA), although their type of engineers is different.
Nonetheless, we leave such a comparison of distances and attributes and the causal analysis to future research.\looseness=-1 


\subsection{Who initiates the ideation contest?}
When using social media for idea generation in practice, it is important to know who to track.
Therefore, understanding the roles and attributes of those who create and spread ideas could help extract ideas from social media more efficiently.\looseness=-1 

To distinguish those who initiate the ideas, we utilize the following network and examine whether the authors of Idea tweets had seen other Idea tweets before they posted theirs. 
Although Twitter API does not provide whether or not a user actually `sees' a certain tweet, we approximate it based on the posted time of the Idea tweet. 
In other words, we assume that a user saw an Idea tweet if any of the followee had posted or shared an Idea tweet before a user's Idea tweet.
Then, we label the authors of Idea tweets as \emph{Primary} if they have not seen any idea tweets beforehand, and \emph{Secondary} if their followees have already tweeted or shared the ideas. 
Among the 2,081 unique authors of Idea tweets (i.e., Primary and Secondary), the number of Primary users was 974 (46.8\%), almost half of the authors.
Also, we label the users who just shared ideas without adding new ideas as \emph{Spreader} (1,831 users). \looseness=-1 

\begin{figure}[!htb]
\centering
\includegraphics[width=0.95\linewidth]{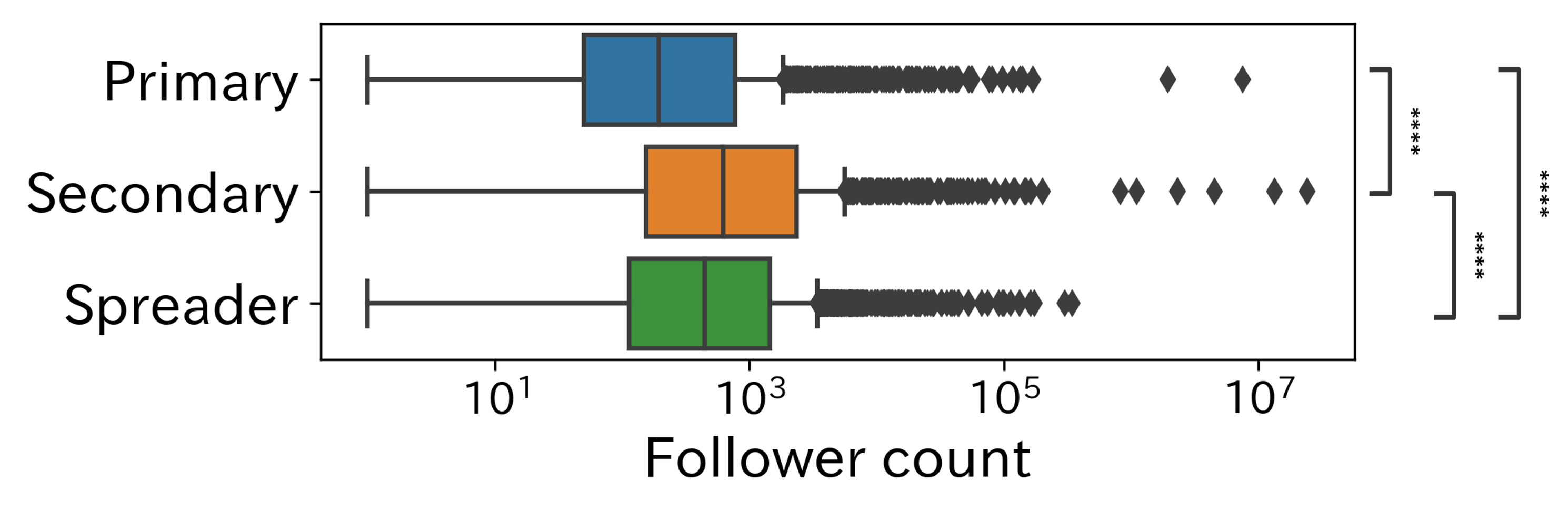}
\caption{Boxplot of follower counts for accounts of Primary, Secondary, and Spreader. The stars indicate the significant differences with Mann-Whitney test with Bonferroni correction (****: $p<$0.0001).}
\label{fig:follower_counts} 
\end{figure}

When we look at the follower counts, the median follower counts of Primary, Secondary, and Spreader were 164, 540, and 446, respectively (shown in Figure~\ref{fig:follower_counts}).
Secondary has the largest number of followers on average, followed by Spreader and Primary, and the differences are all significant ($p<$0.0001 with Mann-Whitney $U$ test with Bonferroni correction).
In other words, it can be seen that the ideas of Primary with fewer followers are picked up by Secondary and Spreader with more followers.
As for Spreader, it has been said that widely spread information is associated with influencers with a large number of followers~\cite{bakshy2011everyone}.
On the other hand, in the comparison between Primary and Secondary, it is interesting to see the former has fewer followers, indicating the difficulty in tracking the users who tend to initiate ideation contests.
Note that the mean (max) follower counts of Primary, Secondary, and Spreader were 5,652 (1.9 mil.), 53,609.4 (24.2 mil.), and 3288.8 (0.8 mil.), respectively.\looseness=-1 

\begin{figure}[!htb]
\centering
\includegraphics[width=0.95\linewidth]{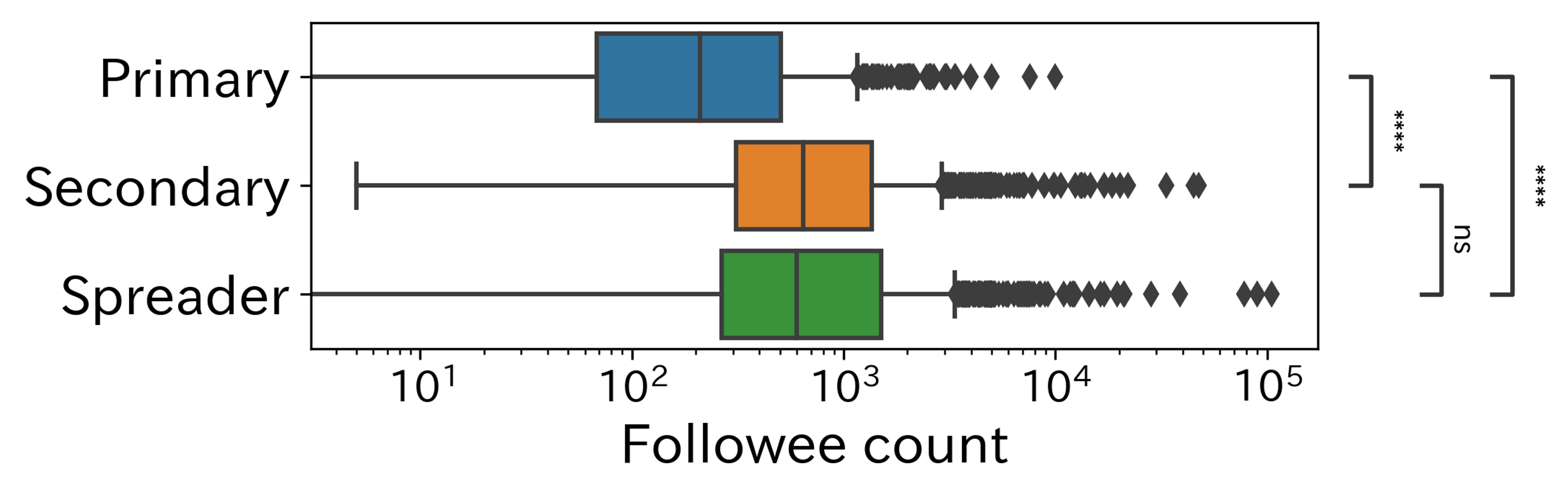}
\caption{Boxplot of followee counts for accounts of Primary, Secondary, and Spreader. 
The stars indicate the significant differences with the Mann-Whitney test with Bonferroni correction (****: $p<$0.0001, ns: $p>0.05$).}
\label{fig:followee_counts} 
\end{figure}

Figure~\ref{fig:followee_counts} also shows the number of followees of these three groups.
The median followee counts of Primary, Secondary, and Spreader were 210, 643, and 602, respectively
Only the Primary group has a significantly lower number of followees, indicating that they tend to lead in proposing original ideas with fewer information sources or that they tend to `come up with' ideas by themselves.
Note that the mean (max) followee counts of Primary, Secondary, and Spreader were 406.3 (9,979), 1,370.8 (47,445), and 1414.4 (105,504), respectively.\looseness=-1 

\section{Idea Exchanges among the Participants}
Ideation contests on social media can be the place of `idea exchanges' through discussions.
In particular, we examine how interactions between users on social media lead to a diversity of ideas. 
On social media, users can see each other's tweets and can exchange ideas by sending replies to others.
It is said that participant interaction in idea generation promotes diversity of ideas~\cite{janis1977decision}, and we examine whether this phenomenon can be observed in social media interactions.\looseness=-1 

Here, we analyze how ideas are exchanged and how discussions help to generate ideas on social media by constructing reply trees.
To do so, we collect all the replies to the 1,766 regular tweets by using the conversation\_id in the Twitter API, as noted in Section $\S$3.1. 
As a result, we obtained 2,096 replies.
The number of regular tweets that received replies is 403 (22.8\%) out of 1,766.
The number of direct replies to the regular tweets varies, ranging from 1 at the minimum and median, 3.93 at the mean, to 64 at the maximum, indicating that some regular tweets received an extremely large number of replies.
Then, we connect the collected replies to form reply trees.
Starting from the 403 regular tweets that received replies, treating a connected chain to the end of replies as one pattern, we form a total of 1,585 patterns of reply trees, including branches from the same crotch.
The maximum length of a reply tree was 10, the minimum 2, the mean 2.40, and the median 2.\looseness=-1

\begin{figure}[!htb]
\centering
\includegraphics[width=0.75\linewidth]{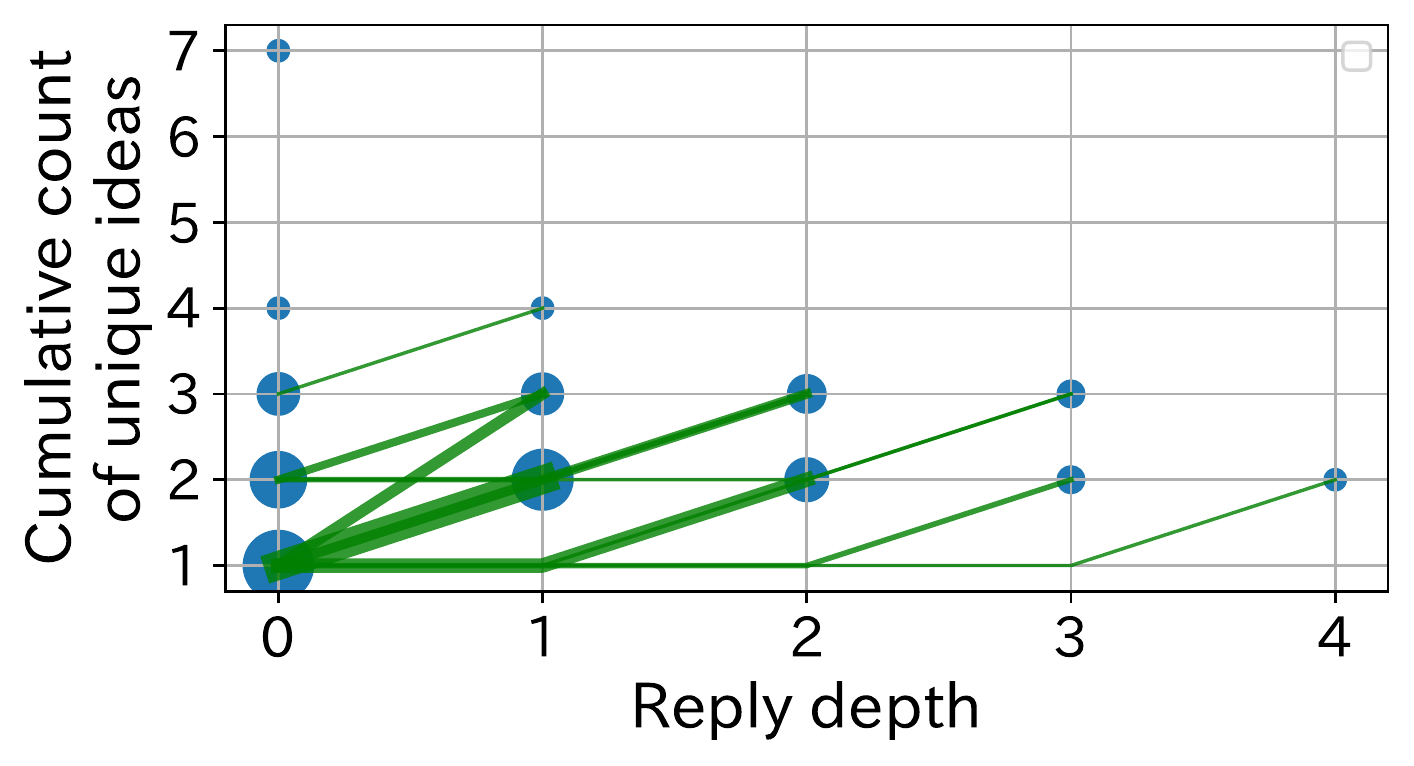}
\caption{The relationship between the depth of reply trees and cumulative count of unique ideas in the tree. Size of points indicates the number of tweets which are lastly added new ideas at a depth of reply (x-axis) and the number of ideas (y-axis). The line indicates the pattern of accumulation of ideas through reply trees, and its width indicates the number of each pattern of reply trees.}
\label{fig:reply_ideas} 
\end{figure}

We first investigate how new ideas are introduced in those reply trees.
We trace the reply tree from the root and count how many unique ideas existed in each reply in the tree.
As a result, among the 1,585 reply trees, we found that 360 trees accumulate new ideas at some point in replies. 
Of these, new ideas are added in the first reply in 327 (90.8\%) trees.
At the maximum, a new idea was added at a reply depth of 4 for the first time.
Figure~\ref{fig:reply_ideas} summarizes the relationship between the depth of replies and the number of ideas.\looseness=-1 

\begin{figure}[!htb]
\centering
\includegraphics[width=0.90\linewidth]{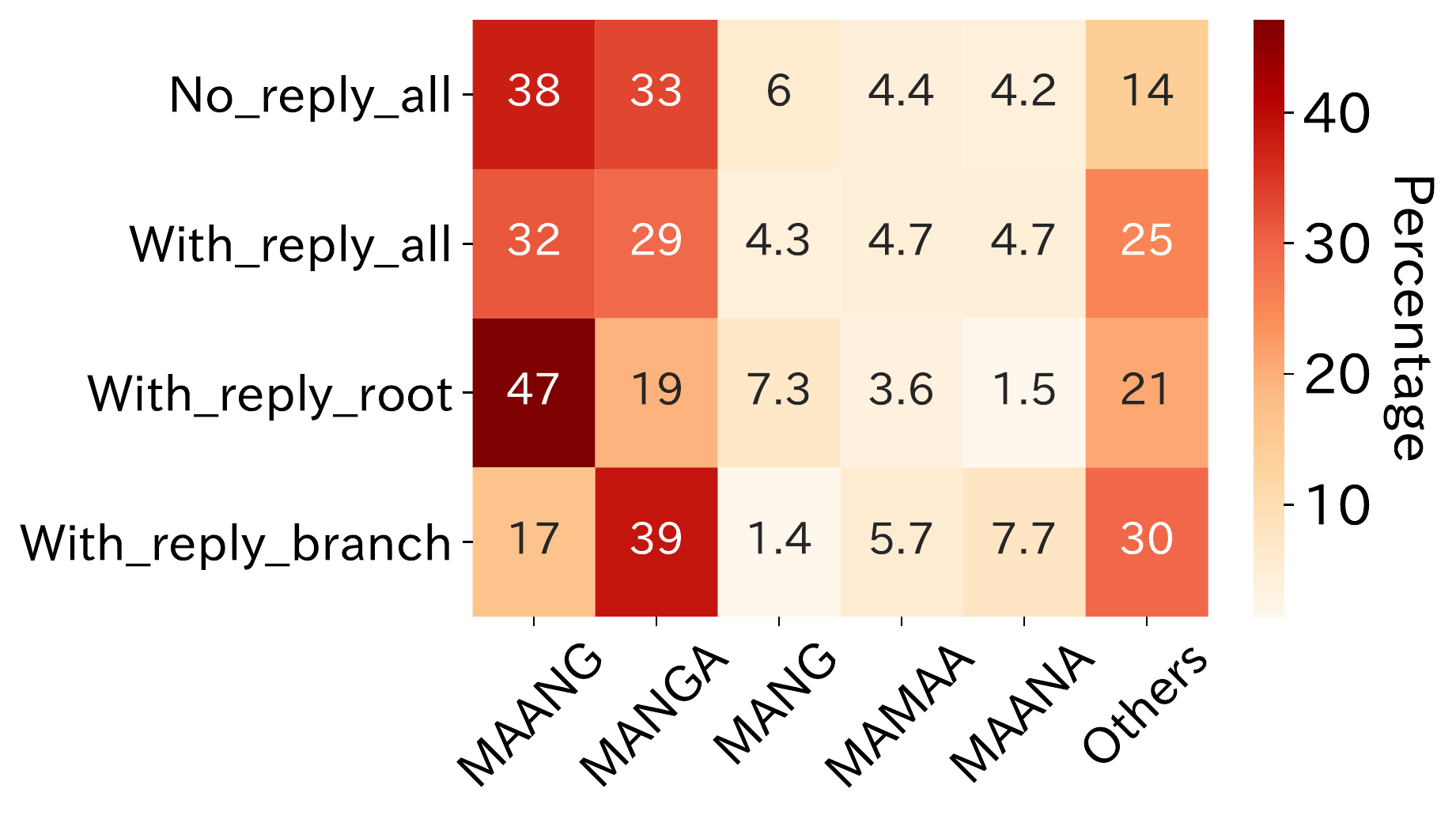}
\caption{
Relationship between type of reply trees and ideas. 
The numbers (aligned with colors) indicate the percentage of ideas posted by each group of users (the sum of row numbers is 100).
}
\label{fig:heatmap3} 
\end{figure}

Next, we analyze whether there is an association between the replies and the diversity of ideas.
To this aim, we tabulated the types of ideas within each chain pattern.
The types of reply trees here were Idea tweets that received no replies (No\_reply\_all), Idea tweets that received replies and all of their replies (With\_reply\_all), only Idea tweets that received replies (With\_reply\_root), and only the replies to the Idea tweets (With\_reply\_branch).
For each, we aggregated the ideas within each group of tweets and created a heatmap of the proportions (Figure~\ref{fig:heatmap3}).
The pair-wise $\chi^2$ tests for all these tweet groups all showed significant differences ($p < 0.001$).
Concerning the presence of replies (No\_reply\_all v.s. With\_reply\_all), the distribution of With\_reply\_all is flatter, and its ratio for Others is larger than No\_reply\_all. 
In other words, a greater variety of ideas was generated when there were replies.
Furthermore, when we look in-depth at With\_reply\_all, we find that MAANG is dominated in the first tweet (With\_reply\_root), while MANGA is dominated in its replies (With\_reply\_branch), and its ratio of Others is also larger. 
In other words, it is conceivable that unordinary ideas are added to the original tweets via replies.\looseness=-1 

Since the diversity of ideas seems to be associated with the interaction of users, it would be helpful to know in which conditions the interaction tends to occur to find them efficiently or spark them deliberately.
Thus, we performed a logistic multiple regression analysis to determine which idea tweets were more likely to receive a reply. 
Here, we constructed a classification model based on whether or not a reply was received at least once (1 if received, 0 otherwise).
We considered the following features:
\textbf{Engineer}: whether the author is an engineer or not. In this case, we used the binary of whether the author belongs to the two clusters of engineers introduced in Section $\S$6.
\textbf{Follower count}: we used the average number of followers of the author.
\textbf{Bio length}: how long it describes itself or how anonymous it is.
\textbf{Verified}: whether or not they have public recognition on Twitter~\cite{TwitterV40:online}.
\textbf{Time}: how long (in seconds) it has been since Facebook changed its name.
\textbf{Negative and positive}: How negative or positive the text is. We used the roBERTa-based pre-trained model~\cite{barbieri2020unified}. This model outputs three classes, including neutral, but neutral was excluded from the regression model.
\textbf{Idea count}: how many ideas an Idea tweet contains.
\textbf{Ideas}: whether a tweet includes the top 5 ideas.\looseness=-1 

Table~\ref{table:regression} shows the results of the logistic regression (number of fake tweets is N = 1,766). 
The p-values are computed using two-tailed z-tests. 
All variance inflation factors (VIFs)~\cite{o2007caution} are less than three, indicating that multicollinearity is negligible. 
Pseudo $R^2$ value is 0.21.
We conduct log transform to some features when needed, indicated as (log) in Table~\ref{table:regression}.
The results show that engineers with more followers and more bio length are more likely to receive replies. 
We also found that the faster one posts an Idea tweet, the more likely it gets replies. 
Interestingly, negative tweets (e.g., ``MAANG sounds a lot less cool than FAANG'') are more likely to receive replies, presumably because negative emotion tends to provoke discussions than positive tweets.\looseness=-1

\begin{table}[!htb]
\centering
\caption{Results of the regression analysis predicting whether or not an Idea tweet gets at least one reply.}
\resizebox{9.0cm}{!}{
\begin{tabular}{p{2.5cm}p{1.2cm}p{0.7cm}p{0.7cm}}
\hline
                     & Coefficient  & Mean  & Std  \\ \hline
const                & -2.01 & -     & -    \\
Engineer             & 0.81${}^{***}$  & 0.22  & 0.42 \\
Follower count (log) & 0.41${}^{***}$  & 5.81  & 2.34 \\
Bio length  (log)    & 0.26${}^{**}$  & 4.13  & 1.26 \\
Verified             & -0.11 & 0.04  & 0.19 \\
Time (log)           & -0.26${}^{***}$ & 11.16 & 1.37 \\
negative             & 1.21${}^{***}$  & 0.17  & 0.20 \\
positive             & -0.29 & 0.22  & 0.24 \\
Idea count           & -0.39 & 1.08  & 0.33 \\
MAANG                & -0.07 & 0.45  & 0.50 \\
MANGA                & 0.22  & 0.33  & 0.47 \\
MANG                 & 0.43  & 0.06  & 0.24 \\
MAMAA                & -0.57 & 0.06  & 0.23 \\
MAANA                & -0.78${}^{*}$ & 0.05  & 0.21 \\ \hline

\multicolumn{4}{l}{
  Significance codes:
  ***$p<0.001$, 
  **$p<0.01$, 
  *$p<0.05$
  }
\end{tabular}
}
\label{table:regression} 
\end{table}


\section{Discussion and Conclusion}

In this study, we used Facebook's name change event as a case study to characterize ideation behavior on social media.
We extracted a comprehensive set of acronyms to refer to new Big Tech companies and analyzed their dynamics and diversity.
The main conclusions are as follows: 1) more diverse ideas are appeared on social media than on mainstream media, 2) the user's topological position on a social network is important for the preference of ideas, and 3) social interactions can spark diversity of ideas.
We also characterize the popularity (e.g., shares and likes) of each idea, those who take different roles in generating ideas, and conditions that idea tweets would attract more replies.
Our study confirms the potential of social media for idea generation and provides a guideline on how to spark diverse ideas.\looseness=-1 

\subsection{Limitation and Future Works}
There are some limitations in our work.
First, we could not track the ideas of private accounts or deleted accounts. 
However, given that the new name of Tech companies is not about a sensitive issue, we believe that their effect would be marginal.
Second, as we examined the single case of the idea contest, more studies are required for the generality of our findings. 
Third, while we examined idea generation behavior on social media, we did not investigate the quality of ideas. 
As \cite{acar2019crowdsourcing} has already pointed out, the ideas gathered by ideation contests are a mixed bag and methods for extracting good ideas from them are in demand.
Fourth, our analysis of the relationship between replies and diversity of ideas does not directly examine causality. 
Future research could study causal inferences from observational data and field studies involving interventions~\cite{mosleh2022field}.\looseness=-1 

We also note that we loosely define an ideation contest as a situation in which many people are exchanging ideas. 
In particular, we did not decide on the winner based on the ideas submitted.
By contrast, in practice, companies may rank the ideas based on the number of RTs or likes, which would eventually become a form of contest.
Also, we have used social media data to characterize and evaluate ideas during the ideation contest, but their impact has not been studied.
For example, MAANG and MANGA, which were the most proposed, did not necessarily replace the original FAANG.
Nonetheless, we believe this study presents meaningful results in terms of characterizing the idea generation mechanism and the competition between new ideas. \looseness=-1

As for the methods for discovering Idea words, a more sophisticated one would be useful for future research. 
In this study, we decided to set a strict search query that considers the context of the Big Tech name change.  
Furthermore, we do not distinguish between suggesting and mentioning an idea.
We considered tweeting using the new Big Tech name to be in itself the same as suggesting it.
However, this can only be valid for this case study.
For other events, there may be a need to extract tweets that are proposing ideas only~\cite{ozcan2021social}.\looseness=-1

As for the volume of data, this study's ideation competition had about 2,000 participants, which is a relatively small size of data compared to common Twitter studies using `Big' data.
This is the result of strict filtering schemes applied to our initial 3.55 million tweets containing candidate terms, and we got excellent precision (97\%: 74 / (2219+74)) from detected idea tweets. 
Nonetheless, considering that it is difficult and expensive to recruit 2,000 people via a corporate platform or crowdsourcing, this study exhibited the usefulness of social media in recruiting such a large number of participants.\looseness=-1 

\bibliographystyle{./IEEEtran}
\bibliography{./IEEEabrv,./main}

\begin{thebibliography}{10}
\providecommand{\url}[1]{#1}
\csname url@samestyle\endcsname
\providecommand{\newblock}{\relax}
\providecommand{\bibinfo}[2]{#2}
\providecommand{\BIBentrySTDinterwordspacing}{\spaceskip=0pt\relax}
\providecommand{\BIBentryALTinterwordstretchfactor}{4}
\providecommand{\BIBentryALTinterwordspacing}{\spaceskip=\fontdimen2\font plus
\BIBentryALTinterwordstretchfactor\fontdimen3\font minus
  \fontdimen4\font\relax}
\providecommand{\BIBforeignlanguage}[2]{{%
\expandafter\ifx\csname l@#1\endcsname\relax
\typeout{** WARNING: IEEEtran.bst: No hyphenation pattern has been}%
\typeout{** loaded for the language `#1'. Using the pattern for}%
\typeout{** the default language instead.}%
\else
\language=\csname l@#1\endcsname
\fi
#2}}
\providecommand{\BIBdecl}{\relax}
\BIBdecl

\bibitem{boeddrich2004ideas}
H.-J. Boeddrich, ``Ideas in the workplace: a new approach towards organizing
  the fuzzy front end of the innovation process,'' \emph{Creativity and
  innovation management}, vol.~13, no.~4, pp. 274--285, 2004.

\bibitem{gatzweiler2017dark}
A.~Gatzweiler, V.~Blazevic, and F.~T. Piller, ``Dark side or bright light:
  Destructive and constructive deviant content in consumer ideation contests,''
  \emph{Journal of Product Innovation Management}, vol.~34, no.~6, pp.
  772--789, 2017.

\bibitem{bayus2013crowdsourcing}
B.~L. Bayus, ``Crowdsourcing new product ideas over time: An analysis of the
  dell ideastorm community,'' \emph{Management science}, vol.~59, no.~1, pp.
  226--244, 2013.

\bibitem{gamber2021effort}
M.~Gamber, T.~Kruft, and A.~Kock, ``Which effort pays off? analyzing
  ideators’ behavioral patterns on corporate ideation platforms,''
  \emph{Journal of Product Innovation Management}, 2021.

\bibitem{Majchrzak2009HarnessingTP}
A.~Majchrzak, L.~Cherbakov, and B.~Ives, ``Harnessing the power of the crowds
  with corporate social networking tools: How ibm does it,'' \emph{MIS Q.
  Executive}, vol.~8, 2009.

\bibitem{muninger2019value}
M.-I. Muninger, W.~Hammedi, and D.~Mahr, ``The value of social media for
  innovation: A capability perspective,'' \emph{Journal of Business Research},
  vol.~95, pp. 116--127, 2019.

\bibitem{piller2012social}
F.~T. Piller, A.~Vossen, and C.~Ihl, ``From social media to social product
  development: the impact of social media on co-creation of innovation,''
  \emph{Die Unternehmung}, vol.~65, no.~1, 2012.

\bibitem{sajid2016social}
S.~Sajid, ``Social media and its role in marketing,'' \emph{Business and
  Economics Journal}, vol.~07, 01 2015.

\bibitem{jeppesen2010marginality}
L.~B. Jeppesen and K.~R. Lakhani, ``Marginality and problem-solving
  effectiveness in broadcast search,'' \emph{Organization science}, vol.~21,
  no.~5, pp. 1016--1033, 2010.

\bibitem{bharati2021idea}
P.~Bharati, K.~Du, A.~Chaudhury, and N.~M. Agrawal, ``Idea co-creation on
  social media platforms: Towards a theory of social ideation,'' \emph{ACM
  SIGMIS Database: the DATABASE for Advances in Information Systems}, vol.~52,
  no.~3, pp. 9--38, 2021.

\bibitem{cao2022visualizing}
Y.~Cao, Y.~Dong, M.~Kim, N.~G. MacLaren, S.~Pandey, S.~D. Dionne, F.~J.
  Yammarino, and H.~Sayama, ``Visualizing collective idea generation and
  innovation processes in social networks,'' \emph{IEEE Transactions on
  Computational Social Systems}, 2022.

\bibitem{han2020computational}
J.~Han, D.~Park, H.~Forbes, and D.~Schaefer, ``A computational approach for
  using social networking platforms to support creative idea generation,''
  \emph{Procedia CIRP}, vol.~91, pp. 382--387, 2020.

\bibitem{FAANGSto80:online}
J.~FERNANDO, ``Faang stocks definition,''
  \url{https://www.investopedia.com/terms/f/faang-stocks.asp}, 2014, (Accessed
  on 02/03/2022).

\bibitem{koh2019adopting}
T.~K. Koh, ``Adopting seekers’ solution exemplars in crowdsourcing ideation
  contests: antecedents and consequences,'' \emph{Information Systems
  Research}, vol.~30, no.~2, pp. 486--506, 2019.

\bibitem{yang2009open}
Y.~Yang, P.-Y. Chen, and P.~Pavlou, ``Open innovation: An empirical study of
  online contests,'' \emph{ICIS 2009 Proceedings}, p.~13, 2009.

\bibitem{deo2021idea}
S.~Deo, A.~Blej, S.~Kirjavainen, and K.~Holtta-Otto, ``Idea generation
  mechanisms: Comparing the influence of classification, combination, building
  on others, and stimulation mechanisms on ideation effectiveness.''
  \emph{Journal of Mechanical Design}, pp. 1--46, 2021.

\bibitem{girotra2010idea}
K.~Girotra, C.~Terwiesch, and K.~T. Ulrich, ``Idea generation and the quality
  of the best idea,'' \emph{Management science}, vol.~56, no.~4, 2010.

\bibitem{van2020idea}
W.~van Osch and B.~Bulgurcu, ``Idea generation in enterprise social media: Open
  versus closed groups and their network structures,'' \emph{Journal of
  Management Information Systems}, vol.~37, no.~4, pp. 904--932, 2020.

\bibitem{bjork2009good}
J.~Bj{\"o}rk and M.~Magnusson, ``Where do good innovation ideas come from?
  exploring the influence of network connectivity on innovation idea quality,''
  \emph{Journal of Product Innovation Management}, vol.~26, 2009.

\bibitem{huizingh2011open}
E.~K. Huizingh, ``Open innovation: State of the art and future perspectives,''
  \emph{Technovation}, vol.~31, no.~1, pp. 2--9, 2011.

\bibitem{ind2013meanings}
N.~Ind and N.~Coates, ``The meanings of co-creation,'' \emph{European business
  review}, 2013.

\bibitem{von2016free}
E.~Von~Hippel, \emph{Free innovation}.\hskip 1em plus 0.5em minus 0.4em\relax
  The MIT Press, 2016.

\bibitem{ihl2012all}
C.~Ihl, A.~Vossen, and F.~T. Piller, ``All for the money? the ambiguity of
  monetary rewards in firm-initiated ideation with users,'' \emph{The Ambiguity
  of Monetary Rewards in Firm-Initiated Ideation with Users}, 2012.

\bibitem{hossain2015ideation}
M.~Hossain and K.~Islam, ``Ideation through online open innovation platform:
  Dell ideastorm,'' \emph{Journal of the Knowledge Economy}, vol.~6, no.~3, pp.
  611--624, 2015.

\bibitem{roberts2016finding}
D.~L. Roberts and F.~T. Piller, ``Finding the right role for social media in
  innovation,'' \emph{MIT Sloan Management Review}, vol.~57, no.~3, 2016.

\bibitem{banks2014social}
\BIBentryALTinterwordspacing
L.~Banks, ``Inspiration meets social media,'' \emph{The New York Times}, 2014.
  [Online]. Available:
  \url{https://www.nytimes.com/2014/12/02/fashion/in-fashion-inspiration-meets-social-media.html}
\BIBentrySTDinterwordspacing

\bibitem{villioti2018social}
\BIBentryALTinterwordspacing
VILLIOTI, ``The influence of social media on fashion design,'' \emph{VILLIOTI},
  2018. [Online]. Available:
  \url{https://www.villiotifashioninstitute.co.za/the-influence-of-social-media-on-fashion-design/}
\BIBentrySTDinterwordspacing

\bibitem{cole2015social}
H.~Cole-Lewis, J.~Pugatch, A.~Sanders, A.~Varghese, S.~Posada, C.~Yun,
  M.~Schwarz, E.~Augustson \emph{et~al.}, ``Social listening: a content
  analysis of e-cigarette discussions on twitter,'' \emph{Journal of medical
  Internet research}, vol.~17, no.~10, p. e4969, 2015.

\bibitem{carr2015social}
J.~Carr, L.~Decreton, W.~Qin, B.~Rojas, T.~Rossochacki, and Y.~wen Yang,
  ``Social media in product development,'' \emph{Food quality and preference},
  vol.~40, pp. 354--364, 2015.

\bibitem{ozcan2021social}
S.~Ozcan, M.~Suloglu, C.~O. Sakar, and S.~Chatufale, ``Social media mining for
  ideation: Identification of sustainable solutions and opinions,''
  \emph{Technovation}, vol. 107, p. 102322, 2021.

\bibitem{cunha2011analyzing}
E.~Cunha, G.~Magno, G.~Comarela, V.~Almeida, M.~A. Gon{\c{c}}alves, and
  F.~Benevenuto, ``Analyzing the dynamic evolution of hashtags on twitter: a
  language-based approach,'' in \emph{Proceedings of the workshop on language
  in social media (LSM 2011)}, 2011, pp. 58--65.

\bibitem{sato2021exploration}
H.~Sato, Y.~Hashimoto, M.~Oka, and T.~Ikegami, ``Exploration in evolutionary
  space by hashtag evolution on a social network service,'' in \emph{ALIFE
  2021: The 2021 Conference on Artificial Life}.\hskip 1em plus 0.5em minus
  0.4em\relax MIT Press, 2021.

\bibitem{dahan2002product}
E.~Dahan and J.~R. Hauser, ``Product development: managing a dispersed
  process,'' \emph{Handbook of marketing}, pp. 179--222, 2002.

\bibitem{janis1977decision}
I.~L. Janis and L.~Mann, \emph{Decision making: A psychological analysis of
  conflict, choice, and commitment.}\hskip 1em plus 0.5em minus 0.4em\relax
  Free press, 1977.

\bibitem{paulus2014creativity}
P.~B. Paulus, T.~S. Larey, and M.~T. Dzindolet, ``Creativity in groups and
  teams,'' in \emph{Groups at work}.\hskip 1em plus 0.5em minus 0.4em\relax
  Psychology Press, 2014, pp. 333--352.

\bibitem{csikszentmihalyi1997flow}
M.~Csikszentmihalyi, ``Flow and the psychology of discovery and invention,''
  \emph{HarperPerennial, New York}, vol.~39, 1997.

\bibitem{inakage2007collective}
M.~Inakage, ``Collective creativity: toward a new paradigm for creative
  culture,'' in \emph{Proceedings of the 2nd International Conference on
  Digital interactive Media in Entertainment and Arts}, 2007, pp. 8--8.

\bibitem{west1996innovation}
M.~A. West and W.~M. Altink, ``Innovation at work: Individual, group,
  organizational, and socio-historical perspectives,'' \emph{European Journal
  of Work and Organizational Psychology}, vol.~5, no.~1, pp. 3--11, 1996.

\bibitem{leenders2003virtuality}
R.~T.~A. Leenders, J.~M. Van~Engelen, and J.~Kratzer, ``Virtuality,
  communication, and new product team creativity: a social network
  perspective,'' \emph{Journal of Engineering and technology management},
  vol.~20, no. 1-2, pp. 69--92, 2003.

\bibitem{dennis1993computer}
A.~R. Dennis and J.~S. Valacich, ``Computer brainstorms: More heads are better
  than one.'' \emph{Journal of applied psychology}, vol.~78, no.~4, 1993.

\bibitem{johnson2021neglect}
D.~R. Johnson, A.~S. Cuthbert, and M.~E. Tynan, ``The neglect of idea diversity
  in creative idea generation and evaluation.'' \emph{Psychology of Aesthetics,
  Creativity, and the Arts}, vol.~15, no.~1, p. 125, 2021.

\bibitem{parjanen2010collective}
S.~Parjanen, V.~Harmaakorpi, and T.~Frantsi, ``Collective creativity and
  brokerage functions in heavily cross-disciplined innovation processes,''
  \emph{Interdisciplinary Journal of Information, Knowledge, and Management},
  vol.~5, p.~1, 2010.

\bibitem{parjanen2012brokerage}
S.~Parjanen, L.~Hennala, and S.~Konsti-Laakso, ``Brokerage functions in a
  virtual idea generation platform: Possibilities for collective creativity?''
  \emph{Innovation}, vol.~14, no.~3, pp. 363--374, 2012.

\bibitem{Enabling63:online}
A.~Tornes, ``Enabling the future of academic research with the twitter api |
  twitter developer platform,''
  \url{https://developer.twitter.com/en/blog/product-news/2021/enabling-the-future-of-academic-research-with-the-twitter-api},
  01 2021, (Accessed on 06/23/2022).

\bibitem{niwattanakul2013using}
S.~Niwattanakul, J.~Singthongchai, E.~Naenudorn, and S.~Wanapu, ``Using of
  jaccard coefficient for keywords similarity,'' in \emph{Proceedings of the
  international multiconference of engineers and computer scientists}, vol.~1,
  no.~6, 2013, pp. 380--384.

\bibitem{MnetAsia52:online}
Wikipedia, ``Mnet asian music awards - wikipedia,''
  \url{https://en.wikipedia.org/wiki/Mnet_Asian_Music_Awards}, (Accessed on
  06/23/2022).

\bibitem{MakeAmer35:online}
------, ``Make america great again - wikipedia,''
  \url{https://en.wikipedia.org/wiki/Make_America_Great_Again}, (Accessed on
  06/23/2022).

\bibitem{xanthopoulos2016hashtag}
P.~Xanthopoulos, O.~P. Panagopoulos, G.~A. Bakamitsos, and E.~Freudmann,
  ``Hashtag hijacking: What it is, why it happens and how to avoid it,''
  \emph{Journal of Digital \& Social Media Marketing}, vol.~3, no.~4, 2016.

\bibitem{MangaWik51:online}
Wikipedia, ``Manga - wikipedia,'' \url{https://en.wikipedia.org/wiki/Manga},
  (Accessed on 06/23/2022).

\bibitem{FANG_FAANG}
J.~Fernando, ``Faang stocks,''
  https://www.investopedia.com/terms/f/faang-stocks.asp, 2022, (Accessed on
  06/20/2022).

\bibitem{Fintwito76:online}
Fintwit, ``Fintwit on twitter: "breaking: Facebook changes its name to 'meta'
  https://t.co/tuodxwtwgj" / twitter,''
  \url{https://twitter.com/fintwit_news/status/1453785540722110466}, 10 2021,
  (Accessed on 06/23/2022).

\bibitem{TwitterV40:online}
Twitter, ``Twitter verification requirements - how to get the blue check,''
  \url{https://help.twitter.com/en/managing-your-account/about-twitter-verified-accounts},
  (Accessed on 08/31/2022).

\bibitem{ChiSquar48:online}
``Chi-square statistic: How to calculate it / distribution - statistics how
  to,''
  \url{https://www.statisticshowto.com/probability-and-statistics/chi-square/},
  (Accessed on 11/14/2022).

\bibitem{Howtouse34:online}
S.~Rodriguez, ``How to use twitter super follows to earn money,''
  \url{https://www.cnbc.com/2021/09/01/how-to-use-twitter-super-follows-to-earn-money.html},
  09 2021, (Accessed on 08/31/2022).

\bibitem{blondel2008fast}
V.~D. Blondel, J.-L. Guillaume, R.~Lambiotte, and E.~Lefebvre, ``Fast unfolding
  of communities in large networks,'' \emph{Journal of statistical mechanics:
  theory and experiment}, vol. 2008, no.~10, p. P10008, 2008.

\bibitem{backstrom2012four}
L.~Backstrom, P.~Boldi, M.~Rosa, J.~Ugander, and S.~Vigna, ``Four degrees of
  separation,'' in \emph{WebSci}, 2012, pp. 33--42.

\bibitem{jacomy2014forceatlas2}
M.~Jacomy, T.~Venturini, S.~Heymann, and M.~Bastian, ``Forceatlas2, a
  continuous graph layout algorithm for handy network visualization designed
  for the gephi software,'' \emph{PloS one}, vol.~9, no.~6, 2014.

\bibitem{bisgin2012study}
H.~Bisgin, N.~Agarwal, and X.~Xu, ``A study of homophily on social media,''
  \emph{World Wide Web}, vol.~15, no.~2, pp. 213--232, 2012.

\bibitem{bakshy2011everyone}
E.~Bakshy, J.~M. Hofman, W.~A. Mason, and D.~J. Watts, ``Everyone's an
  influencer: quantifying influence on twitter,'' in \emph{WSDM}, 2011, pp.
  65--74.

\bibitem{barbieri2020unified}
F.~Barbieri, J.~Camacho-Collados, L.~Espinosa~Anke, and L.~T. Neves, ``Unified
  benchmark and comparative evaluation for tweet classification,''
  \emph{Findings of the Association for Computational Linguistics}, 2020.

\bibitem{o2007caution}
R.~M. O’brien, ``A caution regarding rules of thumb for variance inflation
  factors,'' \emph{Quality \& quantity}, vol.~41, no.~5, pp. 673--690, 2007.

\bibitem{acar2019crowdsourcing}
O.~A. Acar, ``Why crowdsourcing often leads to bad ideas,'' \emph{Harvard
  Business Review}, 2019.

\bibitem{mosleh2022field}
M.~Mosleh, G.~Pennycook, and D.~G. Rand, ``Field experiments on social media,''
  \emph{Current Directions in Psychological Science}, vol.~31, no.~1, pp.
  69--75, 2022.

\end{thebibliography}

\end{document}